\documentclass[aip,graphicx]{revtex4-1}

\usepackage{graphicx}
\usepackage{dcolumn}
\usepackage{bm}
\usepackage[mathlines]{lineno}

\usepackage[utf8]{inputenc}
\usepackage[T1]{fontenc}
\usepackage{mathptmx}
\usepackage{etoolbox}
\usepackage{amssymb}
\usepackage{amsmath}

\def\Xint#1{\mathchoice
   {\XXint\displaystyle\textstyle{#1}}%
   {\XXint\textstyle\scriptstyle{#1}}%
   {\XXint\scriptstyle\scriptscriptstyle{#1}}%
   {\XXint\scriptscriptstyle\scriptscriptstyle{#1}}%
   \!\int}
\def\XXint#1#2#3{{\setbox0=\hbox{$#1{#2#3}{\int}$}
     \vcenter{\hbox{$#2#3$}}\kern-.5\wd0}}

\def\dashint{\Xint-}

\draft 

\begin{document}


\title{Pre-impact dynamics of a droplet impinging on a deformable surface} 



\author{Nathaniel I. J. Henman}
\affiliation{Nanoengineered Systems Lab, Department of Mechanical Engineering, University College London (UCL)}
\affiliation{Department of Mathematics, University College London}

\author{Frank T. Smith} \email{f.smith@ucl.ac.uk}
\affiliation{Department of Mathematics, University College London}

\author{Manish K. Tiwari} \email{m.tiwari@ucl.ac.uk}
\affiliation{Nanoengineered Systems Lab, Department of Mechanical Engineering, University College London (UCL)}
\affiliation{Wellcome/EPSRC Centre for Interventional and Surgical Sciences (WEISS), University College London}


\date{\today}

\begin{abstract}
The non-linear interaction between air and a water droplet just prior to high-speed impingement on a surface is a phenomenon that has been researched extensively and occurs in a number of industrial settings. The role that surface deformation plays in an air cushioned impact of a liquid droplet is considered here. In a two-dimensional framework, assuming small density and viscosity ratios between the air and the liquid, a reduced system of integro-differential equations is derived governing the liquid droplet free-surface shape, the pressure in the thin air film and the deformation of the surface, assuming the effects of surface tension, compressibility and gravity to be negligible. The deformation of the surface is first described in a rather general form, based on previous membrane-type models. The coupled system is then investigated in two cases: a soft viscoelastic case where the substrate stiffness and (viscous) damping are considered and a more general flexible surface where all relevant parameters are retained. Numerical solutions are presented, highlighting a number of key consequences of substrate deformability on the pre-impact phase of droplet impact, such as reduction in pressure buildup, increased air entrapment and considerable delay to touchdown. Connections (including subtle dependence of the size of entrapped air on the droplet velocity, reduced pressure peaks and droplet gliding) with recent experiments and a large deformation analysis are also presented.
\end{abstract}

\pacs{}

\maketitle 


\section{Introduction}
The impact of a droplet on a deformable surface is a commonly occurring event in a number of industrial and natural settings, such as in anti-icing technologies \citep{wang2016}, ink-jet printing \citep{vandam2004} and rain-induced foliar disease transmission \citep{gilet2015}. The use of surface engineering to control droplet impacts \citep{maitra2014a,maitra2014b} can have hugely desirable effects in all the applications mentioned above, among others. The main forms of surface engineering revolve around the use of microstructured roughness or textured surfaces, but the aim of this paper is to examine the influence of another alterable surface property, deformability. There have been a number of experimental studies into droplet impacts with flexible, or soft deformable, substrates, considering both the pre-impact \citep{langley2020,mitra2021} and post-impact behaviour \citep{weisensee2016,vasileiou2016,howland2016}. There has also been some analytical work on the post-impact behaviour \citep{pegg2018,xiong2020,negus2021} and the pre-impact dynamics of droplet settling \cite{poulain2021}, whereas the novelty in our work lies in analysing the pre-impact behaviour of a high-speed droplet impact with a deformable surface. 

When considering droplet impacts, the pre-impact air cushioning is an important feature to consider. The high pressures caused by the thin air layer as the droplet approaches a surface is sufficient enough to significantly deform the droplet before impact. Experimental evidence of this can be seen in \citet{lesser1983} and \citet{liow2001},
as well as in \citet{thoroddsen2005} using high-speed photography. All these studies were on flat rigid surfaces and showed that the free-surface distortion prior to impact was significant enough to entrap a small pocket of air underneath the droplet. This experimental work was further extended by \citet{langley2020} to that of impacts with soft deformable solids and it was found that these impacts entrapped more air than those on rigid surfaces.

A number of theoretical studies consider air cushioning in droplet impacts, in particular \citet{smith2003} who considered a balancing between the forces of an inviscid droplet approaching a rigid wall with a thin, lubricating air layer in-between. This rational viscous-inviscid interaction work was further extended by \citet{hicks2010,hicks2011,hicks2017} for three-dimensional impacts, impacts with liquid layers and other droplet and impacts with porous media, respectively. Also, \citet{purvis2004} considered the effect of surface tension and \citet{mandre2009,mani2010,hicks2013} considered the compressibility of the air.

Liquid-elastic impacts are also the subject of a large number of studies. For example, on an inviscid basis \citet{korobkin2006} studied the impact of a regular wave on an elastic plate and \citet{khabakhpasheva2013} considered a liquid elastic-wedge impact. Similarly, \citet{duchemin2014} investigated the impact of a rigid body on a floating elastic membrane. Of most relevance here are droplet-elastic impacts or droplet impacts with flexible surfaces. \citet{pegg2018} investigated the post-impact interactions of a droplet impact on an elastic plate, where it was assumed that the plate had a relatively high rigidity so that it would vibrate, rather than just be deformed by the impact. They used an axisymmetric Wagner-style model of a droplet impact, which was solved using the method of normal modes. They found that the presence of substrate elasticity acted to slow down the velocity of the advancing contact line and that the induced oscillations of the substrate lead to the onset of splashing. Also using post-impact axisymmetric Wagner theory, \citet{negus2021} recently investigated droplet impact onto a spring-supported plate, where they found solutions for the composite pressure and force on the plate, and provided an excellent comparison to results obtained via direct numerical simulation. \citet{xiong2020} performed numerical simulations of a droplet impacting a flexible surface using a Lattice-Boltzmann method, investigating the effect of bending stiffness on the contact time and wettability of the droplet on the surface. 

There has also been a significant focus on experimental research on droplet impacts with elastic/flexible surfaces, which has motivated the use of flexible elements in surface engineering and microfluidic devices \citep{nguyen2017}. Early work by \citet{pepper2008} examined droplet impact onto elastic membranes of variable tension. They found that by sufficiently lowering the membrane tension, splashing could be suppressed. This work was further complemented by \citet{howland2016} who investigated the impact of ethanol drops on silicone gels of different stiffnesses, where it was found that the stiffness affects the droplet splashing threshold and could also eliminate splashing all together. In both of these works on splashing it was suggested that very early times after, even prior, to impact were critical in the overall outcome of the droplet impact. \citet{weisensee2016} considered droplet impacts with elastic superhydrophobic surfaces and found that the elasticity of the surface was an additional mechanism for reducing the contact time of a bouncing droplet. \citet{vasileiou2016} came to similar conclusions, and \citet{vasileiou2017} found, by investigating impacts with supercooled droplets, that substrate flexibility can improve icephobicity.

The focus of the present study is on an analytical and numerical investigation into the pre-impact behaviour of a droplet impacting a deformable surface. Understanding the pre-impact behaviour of the droplet is vital in understanding the post-impact behaviour. Although in practice droplet impacts are three-dimensional, we will formulate a simplified two-dimensional model and we will use our study to try to gain a qualitative understanding of the effect of surface deformation on various impact quantities, such as touchdown time, contact pressure and air entrapment, which play an important role in understanding the effect on splashing, spreading and wettability of the droplet post-impact (we note a recent very interesting study by \citet{pegg2019} on elastic surfaces which has some overlap with ours). In \ref{goveqns} we formally define the problem and describe an asymptotic analysis which allows us to define a reduced set of governing equations for the droplet free-surface, the pressure in the air film and the shape of the surface as they evolve and interact. We choose to model the deformable surface by the compliant surface model of \citet{carpenter1985} which includes rigidity, tension, stiffness, inertia and damping and so is representative of a number of different surfaces. The aim of this study is to assess the influence of each of these physical parameters. In \ref{visco} and \ref{flexible} we then numerically investigate the solutions, by performing a parametric study of the parameters of the surface. Connections with experiments are then addressed in \ref{expcon}. In \ref{largeflex} we consider the interesting case of relatively large surface deformations, which points to another link with experiments concerning droplet gliding, while \ref{conc} presents the conclusions.

\section{Model formulation and governing equations} \label{goveqns}

Suppose a two-dimensional liquid droplet of radius $R$ approaches a deformable surface with normal velocity $V$. Initially the droplet will be sufficiently far away from the surface that the pressure between the droplet and the surface is constant and the droplet remains circular. Let $(x^*, y^*)$ be the Cartesian coordinates and $t^*$ be time. Then the bottom free surface of the droplet will initially be 
\begin{equation}
f^*(x^*,t^*)=-\sqrt{R^2-{x^*}^2}-Vt^*+R.
\end{equation}
The deformable surface will be denoted $g^*(x^*,t^*)$ and undisturbed will lie on $y^*=0$. Time is measured such that in the absence of air cushioning the droplet would impact the undisturbed deformable surface at $t^*=0$. 

\begin{figure}
    \includegraphics[width=11cm]{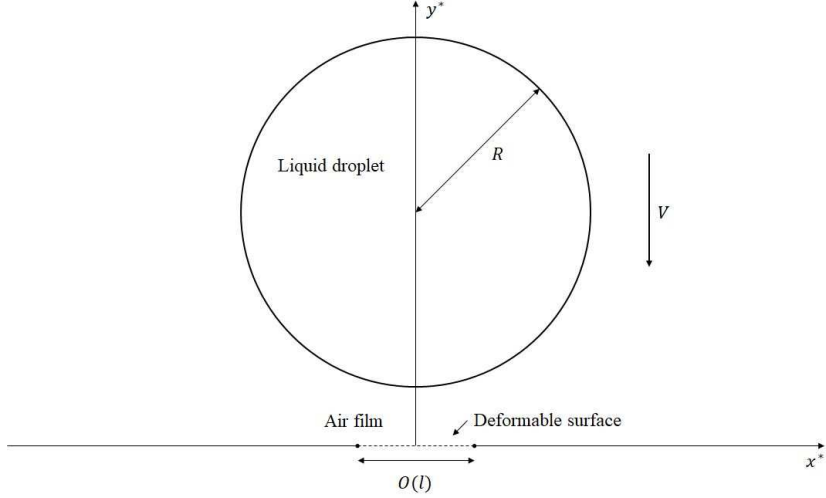}
    \caption{Schematic of the model problem of a two-dimensional droplet of radius $R$ approaching a deformable surface of length $O(l)$ with velocity $V$, with an air film in-between. $l$ is the horizontal length scale over which the interaction of the liquid and air film takes place. The impact is cushioned by the air film between the droplet and the deformable surface.}
    \label{fig:schematic}
\end{figure}

The aim is to derive a system of equations that govern the droplet free-surface, the air film pressure between the droplet and the deformable surface and the shape of the surface. In order to do this the fluid flow will have to be considered separately in the liquid droplet and the air film and an equation governing the shape deformations of the surface will also be considered. These three quantities will form a coupled system. The following derivation will assume that the effects of surface tension, compressibility and gravity are negligible and these assumptions will be discussed in detail later on in this section. The subsequent analysis will exploit the small density ratio $\rho_g\ll\rho_l$, of the liquid ($l$) to the gas ($g$) in order to obtain asymptotically valid equations describing the droplet free-surface, the pressure in the air gap and the shape of the surface. The small quantity used in the asymptotic analysis will be defined as the aspect ratio of the local horizontal length scale $l$, over which the pressure has a leading order effect on the droplet free surface, to the droplet radius $R$,
\begin{equation}
    \varepsilon = \frac{l}{R}.
\end{equation}
Here $l$ is still to be determined. Figure \ref{fig:schematic} shows a schematic of the problem set-up.

All distances are non-dimensionalised with the droplet radius $R$ and time is non-dimensionalised with $R/V$; thus $(x^*, y^*)=R(x, y)$ and $t^*=Rt/V$. Fluid velocity components in both fluids are non-dimensionalised with $V$ and pressure with $\rho_lV^2$. The flow in the droplet and the air film is assumed to be governed by the incompressible Navier-Stokes equations and using the above notation these are
\begin{subequations} 
\label{eq:nsl}
\begin{equation}
\frac{\partial \bm{u}_l}{\partial t}+\bm{u}_l\cdot\bm{\nabla}\bm{u}_l=-\bm{\nabla} p_l+\frac{1}{Re}\nabla^2\bm{u}_l, 
\end{equation}
\begin{equation}
\bm{\nabla}\cdot\bm{u}_l=0, 
\end{equation}
\end{subequations}
for the liquid and
\begin{subequations} 
\label{eq:nsg}
\begin{equation}
\frac{\partial \bm{u}_g}{\partial t}+\bm{u}_g\cdot\bm{\nabla}\bm{u}_g=-\frac{\rho_l}{\rho_g}\bm{\nabla} p_l+\frac{\nu_g}{\nu_l}\frac{1}{Re}\nabla^2\bm{u}_g,
\end{equation}
\begin{equation}
\bm{\nabla}\cdot\bm{u}_g=0
\end{equation}
\end{subequations}
for the air, where $\bm{\nabla}=(\partial/\partial x,\partial/\partial y)$, $\nu_\alpha=\mu_\alpha/\rho_\alpha$ is the kinematic viscosity of the liquid ($\alpha =l$) or gas ($\alpha = g$), $\bm{u}_\alpha=(u_\alpha,v_\alpha)$ is the fluid velocity field and $Re=RV/\nu_l$ is the global Reynolds number based on the properties and initial velocity of the liquid. For the parameter regime of interest, the Reynolds number $Re$ is typically large and this will, again, be discussed at the end of this section.

The non-dimensional initial liquid free surface profile is now given by 
\begin{equation} \label{eq:ndd}
    f(x,t) = -\sqrt{1-x^2}-t+1.
\end{equation}
To close our system we must satisfy the kinematic boundary conditions on the liquid-gas interface and the gas-surface interface,
\begin{equation} \label{eq:kclg}
\frac{\partial f}{\partial t}+u_\alpha\frac{\partial f}{\partial x}=v_\alpha,~~~~~\mbox{at\ } y=f(x,t)
\end{equation}
and
\begin{equation} \label{eq:kcgs}
\frac{\partial g}{\partial t}+u_\alpha\frac{\partial g}{\partial x}=v_\alpha,~~~~~\mbox{at\ } y=g(x,t)
\end{equation}
for $\alpha=l$ or $g$. Neglecting surface tension effects gives the normal stress balance on the liquid-gas interface as
\begin{equation} \label{eq:nsc}
    p_l=p_g.
\end{equation}

\subsection{Liquid droplet}

To determine the behaviour of the liquid droplet free surface, close to the point of initial contact which is $x=y=0$, we take the following scaling
\begin{equation} \label{eq:scalesl}
    (x,y,t,f) = (\varepsilon X, \varepsilon Y, \varepsilon^2 T, \varepsilon^2 F),
\end{equation}
where the scales of $t$ and $f$ come from the desire to study short time behaviour and from the form of equation (\ref{eq:ndd}), respectively. The $O(\varepsilon^2)$ time scale is that expected for the traversing, at a unit non-dimensional velocity, of the thin air gap which has normal width of $O(\varepsilon^2)$. The scales (\ref{eq:scalesl}) lead us to take asymptotic expansions of the liquid velocity components and pressure in the following form
\begin{equation} \label{eq:expanl}
    (u_l,v_l,p_l) = (U_l,V_l,\varepsilon^{-1}P_l)+\cdots.
\end{equation}
The vertical velocity scale is order unity due to the order unity downward velocity of the droplet, then the horizontal velocity scale follows from continuity. The large pressure scale arises from seeking a balance between the liquid acceleration and the pressure gradient.

Now, for $Re\gg1$, upon substitution of scales (\ref{eq:scalesl}) and expansions (\ref{eq:expanl}) into the governing equations (\ref{eq:nsl}), the leading order momentum equations and continuity equation are those of unsteady potential flow
\begin{subequations} \label{eq:potflow}
\begin{equation}
\frac{\partial U_l}{\partial T}=-\frac{\partial P_l}{\partial X},
\end{equation}
\begin{equation}
\frac{\partial V_l}{\partial T}=-\frac{\partial P_l}{\partial Y},
\end{equation}
\begin{equation}
\frac{\partial U_l}{\partial X}+\frac{\partial V_l}{\partial Y} = 0.
\end{equation}
\end{subequations}
The leading order kinematic condition (\ref{eq:kclg}) now reduces to 
\begin{equation} \label{eq:kinl}
    V_l\rightarrow\frac{\partial F}{\partial T},~~~~~\mbox{as\ }Y\rightarrow 0^+,
\end{equation}
while the far-field droplet behaviour reduces to 
\begin{equation} \label{eq:farfieldl}
F(X,T) \sim \frac{X^2}{2}-T+O(\varepsilon),~~~~~\mbox{as\ }|X|\rightarrow\infty~\mbox{or\ }T\rightarrow -\infty,
\end{equation}
from (\ref{eq:ndd}).

From equations (\ref{eq:potflow}) it can be shown that $P_l$ satisfies Laplace's equation and due to the far-field boundedness (\ref{eq:farfieldl}) and condition (\ref{eq:kinl}) the pressure profile on the droplet interface $P(X,T) = P_l(X,0,T)$ and free-surface profile $F(X,T)$ are related by 
\begin{equation} \label{eq:cauchy}
    \frac{\partial^2 F}{\partial T^2}=\frac{1}{\pi}\dashint_{-\infty}^{\infty}\frac{\partial P}{\partial \zeta}(\zeta,T)\frac{\mathrm{d}\zeta}{X-\zeta}.
\end{equation}
The bar denotes the Cauchy principle value integral.

\subsection{Gas cushioning}

In the thin gas film between the droplet and the deformable surface, we assume that the vertical length scale is an order of magnitude smaller than the horizontal length scale to capture the slenderness of the film. Also, unlike in the droplet formulation, we now need to consider the deformable surface $g$, which we suppose to have the same scale as the droplet free surface $f$ in order for the surface deformation to have a leading order influence. We then take the following scaling in terms of our previously defined small parameter $\varepsilon$,
\begin{equation} \label{eq:scalesgas}
    (x,y,t,f,g) = (\varepsilon X,\varepsilon^2 Y,\varepsilon^2 T, \varepsilon ^2 F, \varepsilon^2 G).
\end{equation}
The scales (\ref{eq:scalesgas}) lead to asymptotic expansions of the form
\begin{equation} \label{eq:expgas}
    (u_g,v_g,p_g)=(\varepsilon^{-1}U_g,V_g,\varepsilon^{-1}P_g)+\cdots,
\end{equation}
where the horizontal velocity scale is expected to be large compared with the vertical velocity scale in order to balance the continuity equation. 

When substituting scales (\ref{eq:scalesgas}) and expansions (\ref{eq:expgas}) into the governing equations for the gas (\ref{eq:nsg}), for $Re\gg 1$ the leading order equations take the form of lubrication flow,
\begin{subequations} \label{eq:lubens}
\begin{equation} \label{eq:lubensx}
0=\frac{\partial P_g}{\partial X}+\frac{\partial^2 U_g}{\partial Y^2},
\end{equation}
\begin{equation} \label{eq:lubensy}
0=-\frac{\partial P_g}{\partial Y},
\end{equation}
\begin{equation} \label{eq:lubenscont}
\frac{\partial U_g}{\partial X}+\frac{\partial V_g}{\partial Y}=0.
\end{equation}
\end{subequations}
The inertial and acceleration terms do not appear in the left hand side of equations (\ref{eq:lubensx}-\ref{eq:lubensy}) because these terms are negligible compared to the pressure gradient term. This assumption requires $\rho_g/\rho_l\ll\varepsilon$. Also, it is assumed that there is a balance in the horizontal momentum equation between the pressure gradient and the viscous terms \citep{smith2003}, which leads to the definition
\begin{equation}
    \varepsilon=\left(\frac{\mu_g}{\mu_lRe}\right)^{1/3}.
\end{equation}
This, combined with the requirement $Re\gg1$ in the liquid, gives the condition that $\varepsilon$ must satisfy in order for our model to be valid, namely
\begin{equation}
    \frac{\rho_g}{\rho_l}\ll\varepsilon\ll\left(\frac{\mu_g}{\mu_l}\right)^{1/3}.
\end{equation}
For example, water and air give rise to the range $10^{-3}\ll\varepsilon\ll0.27$.

The leading order kinematic conditions (\ref{eq:kclg}) and (\ref{eq:kcgs}) now reduce to 
\begin{equation} \label{eq:kcsf}
    V_g=\frac{\partial F}{\partial T},~~~~~\mbox{at\ }Y=F(X,T)
\end{equation}
and
\begin{equation} \label{eq:kcsg}
    V_g=\frac{\partial G}{\partial T},~~~~~\mbox{at\ }Y=G(X,T)
\end{equation}
respectively. The vertical momentum equation (\ref{eq:lubensy}) implies that $P_g(X,Y,T)=P_g(X,0,T)$, then integrating equation (\ref{eq:lubensx}) twice in $Y$ and applying conditions (\ref{eq:kcsf}-\ref{eq:kcsg}), with normal stress condition (\ref{eq:nsc}), gives
\begin{equation} \label{eq:lube}
    \frac{\partial}{\partial X}\left(\left(F-G\right)^3\frac{\partial P}{\partial X}\right)=12\frac{\partial}{\partial T}\left(F-G\right),
\end{equation}
the (Reynolds) lubrication equation that helps to link all three unknown quantities $F$, $G$ and $P$.

\subsection{The deformable surface}

Finally, we require another equation linking the pressure in the air film with the deformation of the surface. The model we will use here was first described by \citet{carpenter1985} to model surface coatings as a elastic plate (or tensioned membrane) supported above a rigid surface by an array of springs. It is derived from a nonlinear model that includes all relevant physics which is then linearised under the assumption of longitudinal deflections being much smaller than transverse ones. An excellent derivation of this equation can be found in \citet{alexander2020} (we will ignore viscous traction in the air film in our model). This model is also used in a number of other studies on deformable surfaces \citep{carpenter1986,gajjar1996,davies1997,pruessner2013,pruessner2015}. The relevant equation takes the non-dimensional form
\begin{eqnarray} \label{eq:flexnd}
    e_1\frac{\partial^4 g}{\partial x^4}+e_2\frac{\partial^2 g}{\partial x^2}+e_3g+&&e_4\frac{\partial ^2g}{\partial t^2}+e_5\frac{\partial g}{\partial t} \nonumber\\ 
    &&= p(x,g,t)-p_s(x,g,t),
\end{eqnarray}
where the non-dimensional coefficients $e_i$ are $(e_1,e_2,e_3,e_4,e_5)=(-B^*/R^3V^2\rho_l ,\allowbreak T_t^*/RV^2\rho_l ,\allowbreak -\kappa^*R/V^2\rho_l ,\allowbreak -M^*/R\rho_l ,\allowbreak -C^*/V\rho_l)$ and $p_s$ is the non-dimensional relative base pressure, which is taken to be zero. We choose to use this thin-membrane type model over more simpler models as there is commonality in such membranes in nature and practical settings, such as droplet impact onto leaves \cite{gilet2015}, butterflies \cite{li2018}, umbrellas and raincoats, and relevance in many other applications of droplet impact on deformable surfaces discussed elsewhere in this paper. More simple surface deformation models can be extracted as subsets from equation (\ref{eq:flex}) and one such, a Kelvin-Voigt model of viscoelasticity, is described in detail in \ref{visco} while the equation is considered in full generality in \ref{flexible}. The recent study by \citet{pegg2019} is likewise based on the \citet{smith2003} theory but with a surface equation generally different from ours; the present work is more focused on soft deformable surfaces which are found to lead to new non-intuitive results. The constants $B^*$, $T_t^*$, $\kappa^*$, $M^*$ and $C^*$ correspond to the flexural rigidity, the longitudinal tension, the stiffness, the mass density and the damping constant of the surface, respectively. In order to apply equation (\ref{eq:flexnd}) to our scaled liquid droplet application described above, we must apply scales for $g$, $x$, $t$ and $p$ given in (\ref{eq:scalesgas}-\ref{eq:expgas}). This gives
\begin{equation} \label{eq:flex}
\tilde{e}_1\frac{\partial ^4 G}{\partial X^4}+\tilde{e}_2\frac{\partial^2 G}{\partial X^2}+\tilde{e}_3G+\tilde{e}_4\frac{\partial ^2G}{\partial T^2}+\tilde{e}_5\frac{\partial G}{\partial T}=P-P_s,
\end{equation}
where $(\tilde{e}_1,\tilde{e}_2,\tilde{e}_3,\tilde{e}_4,\tilde{e}_5)=(\varepsilon^{-1}e_1,\varepsilon e_2,\varepsilon^3e_3,\varepsilon^{-1}e_4,\varepsilon e_5)$ and $P_s=\varepsilon^{-1}p_s$. Therefore, the five non-dimensional parameters which control the system are the following
\begin{equation} \label{eq:parameters}
\begin{split}
    &\tilde{e}_1 = -\frac{B^*}{(\rho_l^2\mu_gV^5R^8)^{1/3}},~~~~~\tilde{e}_2=\frac{T_t^*}{(\rho_l^4\mu_g^{-1}V^7R^4)^{1/3}},\\
    &\tilde{e}_3 = -\frac{\kappa^*}{\rho_l^2\mu_g^{-1}V^3}, ~~~~~\tilde{e}_4 = -\frac{M^*}{(\rho_l^2\mu_gV^{-1}R^2)^{1/3}},\\
    &\tilde{e}_5=-\frac{C^*}{(\rho_l^4\mu_g^{-1}V^4R)^{1/3}}.
    \end{split}
\end{equation}
The relative size and importance of each term will be discussed in each case considered in this study. In practise, each of the dimensional structural parameters $B^*$, $T_t^*$, $\kappa^*$, $M^*$ and $C^*$ can vary dramatically depending on the situation, so therefore so can each $e_i$ in (\ref{eq:parameters}). As droplet impact is a ubiquitous phenomenon, it can occur on many different surfaces of differing elastic properties, for example leaves \citep{gilet2015}, thin foils \citep{contino2019}, skin \citep{tagawa2013} and foodstuffs \citep{andrade2012}. It is also possible to use surface engineering to modify the elastic properties of a surface to our advantage \citep{howland2016,weisensee2016,vasileiou2016,vasileiou2017,langley2020}. Therefore our focus will be on studying different cases of equation (\ref{eq:flex}), with a very wide range of value of the parameters $e_i$. 

The deformable surface will be considered to be initially zero and stationary. Boundary conditions will be discussed separately for each case considered. The system to be solved is the non-linear set of governing equations (\ref{eq:cauchy}), (\ref{eq:lube}) and (\ref{eq:flex}), subject to the boundary condition (\ref{eq:farfieldl}), the pressure $P$ being initially zero and decaying at infinity and the relevant boundary and initial condition on $G$. The governing equations require a numerical treatment. The numerical scheme is slightly different for each case considered and so will be discussed separately for each case.

\subsection{Fluid parameter regime and model validity}

It is useful now to mention the fluid parameters used in recent experiments. Of most relevance here, we will consider the experiments performed by \citet{langley2020} for the pre-impact behaviour of ethanol droplets impacting soft surfaces of varying stiffness. They considered a parameter regime where the Reynolds number $Re$ ranged from 1209 to 20394 and the Weber number $We$ ranged from 17 to 2825. Hence both are typically large and in particular the assumption of a quasi-inviscid liquid droplet seems valid in this regime. Also, in the present study surface tension has been ignored and the scaled surface tension forces are given by
\begin{equation}
    \frac{\varepsilon^2}{\rho_lV^2l}\sigma\nabla^2F=\frac{\varepsilon}{We}\nabla^2F,
\end{equation}
where $\sigma$ is the surface tension coefficient and $We=\rho_lV^2R/\sigma$ is the Weber number. As the parameter $\varepsilon$ is small and $We$ is typically large, this again seems a valid assumption for the vast majority of the evolution. As the droplet approaches the surface, we do still expect the free-surface to deform to the point where high curvatures are observed, as seen in \citet{smith2003}. This would result in a large value for $\nabla^2F$, and thus surface tension forces may become significant at this stage. However, for some cases in the following section the free-surface curvature will not become large at any stage and so surface tension effects remain small. The effect of surface tension is considered in \citet{purvis2004} for droplet impacts in our parameter regime and in \citet{vandenbroeck2008} for a similar air-cushioning related problem for higher Reynolds numbers than in this work. In the present study, however, as has already been assumed, surface tension effects will be ignored. For all parameters considered in \citet{langley2020} the Froude number $Fr=V/\sqrt{gR}$, where $g$ is the gravitational acceleration, is also large; hence gravitational effects are ignored. 

The effects of compressibility in the air film are also ignored in the present study. A scaling argument performed by \citet{mandre2009}, where they balanced the gas pressure gradient with the droplet deceleration, found that compressibility effects in the air can be ignored if the compressiblity parameter $\delta$ satisfies
\begin{equation} \label{eq:comp}
    \delta = \frac{p_0^*}{(R\mu_g^{-1}V^7\rho_l^4)^{1/3}}\gg1,
\end{equation}
where $p_0^*$ is the surrounding ambient pressure. In the study of \citet{langley2020} there are impacts considered both in the compressible ($\delta<1$) and incompressible regimes ($\delta\gg1$), so we will still attempt to draw comparisons with their work later in \ref{expcon}. For detailed discussions of compressible gas-cushioned droplet impacts see \citet{mani2010} and \citet{hicks2013} also.

\begin{figure}
    \includegraphics[width=11cm]{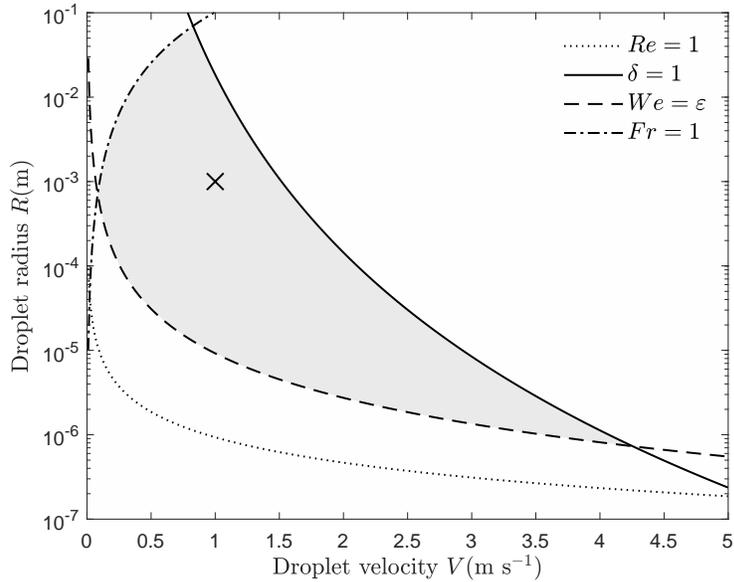}
    \caption{Range of parameter validity for a water droplet impact in air, with $p_0^*=10^5$ Pa. Above the dotted line $Re\gg1$ and the liquid droplet may be considered quasi-inviscid, below the solid line $\delta\ll1$ and compressibility effects in the air may be ignored, above the dashed line $We\gg\varepsilon$ and surface tension forces may be ignored for the vast majority of the evolution, and below the dash-dotted line $Fr\gg1$ and gravitational effects may be ignored. These limits are represented by the gray shaded region. The cross is the point where the droplet radius is 1 mm and the droplet velocity is 1 m s$^{-1}$, a set of values well within our regime and used commonly in this paper for dimensional calculations.}
    \label{fig:parareg}
\end{figure}

Figure \ref{fig:parareg} summarises the main model assumptions for a water droplet impact, when $p_0^*=10^5$ Pa.
There is a clear and rather large range of droplet velocities and radii where our model is valid and this is the gray shaded region in figure \ref{fig:parareg}, where the cross shows where the droplet radius is 1 mm and the velocity is 1 m s$^{-1}$, a point well within our regime.

\section{Impact onto a viscoelastic solid} \label{visco}

The first case that will be considered is that of an air cushioned droplet impact onto a soft viscoelastic solid. The viscoelastic material will be assumed to exhibit no bending, tension or inertia, such as a coating on a rigid surface of infinite extent. In this case we may remove the influence of coefficients $\tilde{e}_1$, $\tilde{e}_2$ and $\tilde{e}_4$ and reduce our deformable surface equation (\ref{eq:flex}) to 
\begin{equation} \label{eq:ve}
    \tilde{e}_3G+\tilde{e}_5\frac{\partial G}{\partial T} = P,
\end{equation}
which gives the simple Kelvin-Voigt model of a solid deforming in reaction to an applied pressure. Droplet impact and dynamics on soft viscoelastic solids has recieved attention both experimentally \cite{chen2011,chen2016} and analytically \cite{leong2020}.

If the soft viscoelastic solid is a coating sitting on an otherwise rigid substrate, the coating depth would need to be $O(\varepsilon^2 R)$ in order for the depth to have an influence on the dynamics. If $d^*$ is the dimensional depth of the substrate, then this can be envisaged as $d^*\ll\varepsilon R$, which for a droplet of radius 1 mm and velocity 1 m s$^{-1}$ becomes $d^*\ll 26~\mu$m. In the experiments of \citet{howland2016} they predominately considered soft substrates of depth 1 cm, well outside this limit, however they did also discuss the effect of a substrate of depth 3 $\mu$m, which is within this limit. It was found that the splashing behaviour of the impacting droplet on this very shallow soft silicone substrate was almost identical to that of a much deeper substrate of hard acrylic. In light of this fairly uninteresting behaviour for substrates of shallow depth, we will limit our study to the case of substrates of apparent infinite depth with zero under-pressure ($P_s=0$). The smallest depth of substrate considered in \citet{langley2020} was 1 mm, also well outside this limit.

Given the form of equation (\ref{eq:ve}) it follows naturally that if the pressure is decaying at infinity, then so is the surface deformation. Therefore the boundary condition at infinity will be $G\rightarrow0$. The viscoelastic model equation (\ref{eq:ve}) is solved in conjunction with the free-surface equation (\ref{eq:cauchy}) and the lubrication equation (\ref{eq:lube}). The numerical method proceeds by first substituting equation (\ref{eq:ve}) into equations (\ref{eq:cauchy}) and (\ref{eq:lube}) for $P$. At each time step, the lubrication equation $(\ref{eq:lube})$ is solved via a finite difference discretization for $G$, which is then used to solve the free-surface equation (\ref{eq:cauchy}) for $F$, using Fast Fourier Transform algorithms. This process is repeated until successive iterates are within a convergence criteria, after which the solution proceeds to the next time step. Once we have a converged solution for $G$, the pressure $P$ is then readily calculated from equation $(\ref{eq:ve})$. This method is akin to that used in \citet{hicks2017}. By eliminating the pressure from our simultaneous equations, the efficiency of the code is improved somewhat as not only do we have two equations to solve as opposed to three, we are no longer having to resolve a diverging pressure solution \citep{smith2003} from the lubrication equation, which can be computationally expensive. Tests were run to ensure the solutions were independent of grid size and time step size. A suitable spatial domain size here was found to be $X\in[-20,20]$, with the boundaries here being non-invasive and the solutions remaining unchanged for further increases to the spatial domain size. The simulations are very sensitive to the chosen start time. We performed a test on the surface with parameters $(\tilde{e}_1,\tilde{e}_2,\tilde{e}_3,\tilde{e}_4,\tilde{e}_5)=(0,0,-0.1,0,-0.1)$ (the most deformable surface considered in this study) with start times of $T=-50$ and $T=-100$. It was found that the difference in the size of entrapped air upon touchdown (defined in \ref{air}) was 2.5\% for these two start times, and so $T=-50$ was chosen as the start time for all simulations.

\subsection{Free-surface and pressure profile evolution}

\begin{figure*} 
    \centering
    \includegraphics[width=15cm]{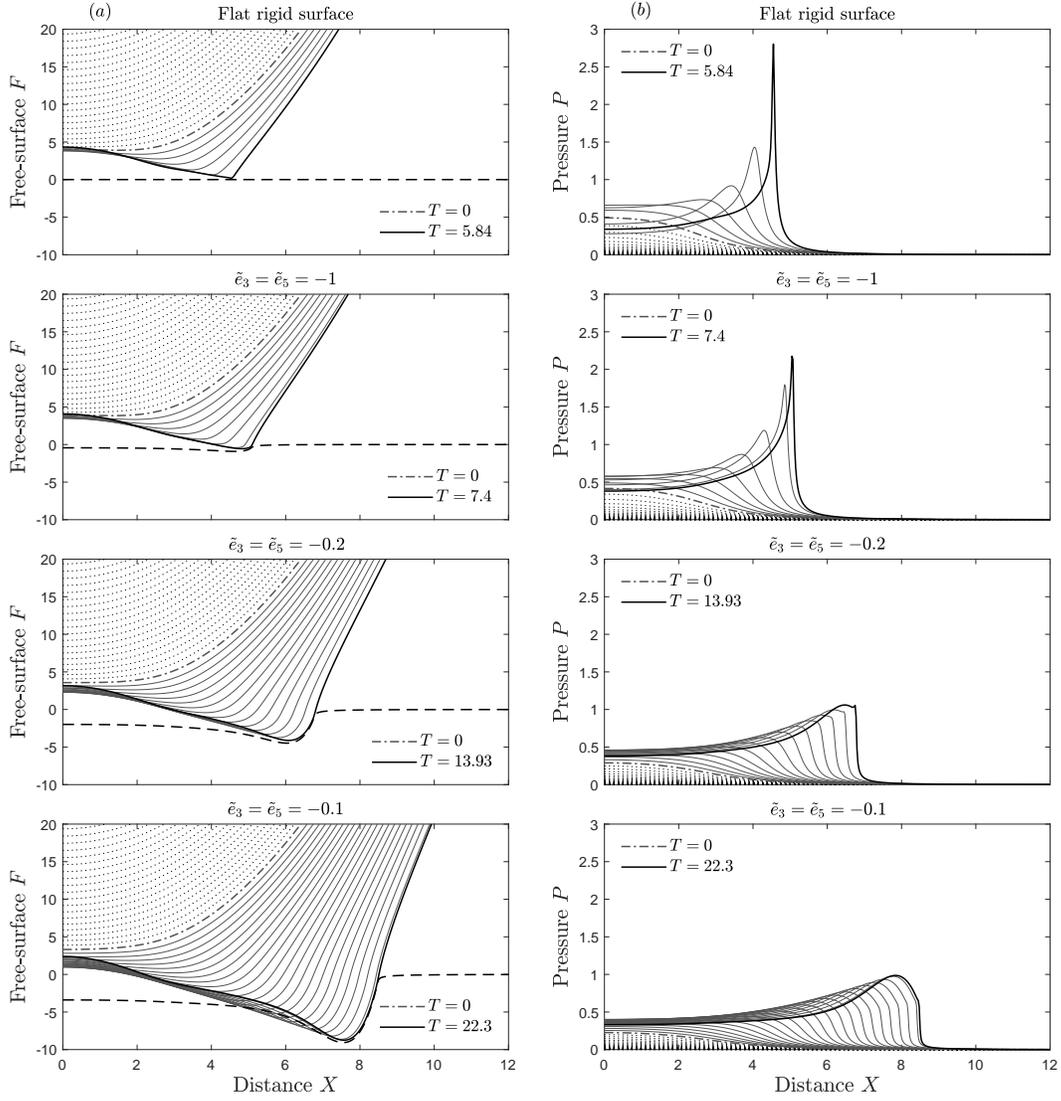}
    \caption{Solution profiles, showing evolution of (\textit{a}) the free-surface height $F$ and (\textit{b}) the pressure $P$ for normal impact of a droplet on a flat rigid surface and deformable viscoelastic surfaces with $\tilde{e}_3=\tilde{e}_5=-1$, $-0.2$ and $-0.1$. The solutions are shown in integer time increments except for the final thick solid line, which is the solution just prior to touchdown, and the dashed line, which is the deformable surface solution $G$ just prior to touchdown. The dash-dotted line is the solution at $T=0$ where the droplet would touchdown on an undisturbed surface in the absence of air cushioning.}
    \label{fig:FGPfullevve}
\end{figure*}

Figure \ref{fig:FGPfullevve} shows the evolution of the droplet free-surface and pressure profile for a flat rigid surface ($G=0$) and for a viscoelastic surface for a range of values of $\tilde{e}_3=\tilde{e}_5$. We chose to vary the parameters as $\tilde{e}_3=\tilde{e}_5$ because in practise they are dependant on each other, however different combinations are considered later on. The top two panels are the solution for a flat rigid surface which has been previously reported a number of times \citep{smith2003,hicks2017}. As the droplet approaches the surface, there is an initial build-up of pressure directly beneath the droplet. The build-up of pressure acts to decelerate the falling droplet and as the gap between the droplet and the surface narrows, the pressure build-up is large enough to decelerate the droplet free-surface to rest at the center point. At this stage, the droplet begins to deform either side of the center point and eventually overtake it, resulting in an approach to touchdown in two locations. This, in turn, results in the pressure bifurcating away from the center point into two pressure peaks. The process continues until the moment of touchdown on the surface, which results in the entrapment of an air pocket, which may then subsequently form a bubble \citep{vandam2004,thoroddsen2005}.

The result for the flat rigid surface in figure \ref{fig:FGPfullevve} is then compared to the result for impact onto a viscoelastic surface for decreasing magnitudes of the surface parameters $\tilde{e}_3=\tilde{e}_5$. Lowering the magnitude of the surface parameters $\tilde{e}_3$ and $\tilde{e}_5$ corresponds to lowering the surface stiffness and (viscous) damping, which results in larger surface deformations. Rows two to four in figure \ref{fig:FGPfullevve} show the solutions for the free-surface and the pressure profile evolution, along with the shape of the surface as touchdown is approached, for $\tilde{e}_3=\tilde{e}_5=-1$, -0.2 and -0.1, respectively. For a water droplet in air of radius 1 mm and impact velocity 1 m s$^{-1}$, these parameter variations correspond to a surface spring stiffness in the range $\kappa^*\sim 10^9$ - $10^{10}$ Pa m$^{-1}$ and (viscous) damping of $C^*\sim 10^3$ - $10^{4}$ Pa s. We note that the spring stiffness values here are larger than the stiffnesses of the soft solids considered in \citet{langley2020}, however the spring stiffness defined in equation (\ref{eq:flex}) is a stiffness $per~unit~length$ and the length scales in this problem are typically very small. Smaller values of the dimensionless parameter $\tilde{e}_3$ are considered in \ref{air}. 

A number of conclusions can be drawn from the influence of surface deformability in this viscoelastic model on the pre-impact dynamics of droplet impact. First of all, there is a profound delay to touchdown. In the absence of air cushioning the droplet would impact on an undisturbed surface at $X=Y=0$ at $T=0$. The presence of air cushioning delays this touchdown into positive time. In figure \ref{fig:FGPfullevve} the solution at $T=0$ is the dash-dotted line, while the positive time solutions are the solid lines. As the droplet approaches the soft viscoelastic surface, the build-up of pressure beneath the droplet acts now to not only deform and decelerate the droplet free surface, but also push the surface away from the droplet. This results in a slower closing of the air gap between the droplet and the soft viscoelastic surface, with lower magnitudes of surface parameters $\tilde{e}_3$ and $\tilde{e}_5$ resulting in a slower closing of the air gap. In consequence there is a larger delay to touchdown as illustrated by the thick solid lines in figure \ref{fig:FGPfullevve}, corresponding to the solution as touchdown is approached, where the time has increased from $T=5.84$ for a flat rigid surface to $T=22.3$ for the soft viscoelastic surface with $\tilde{e}_3=\tilde{e}_5=-0.1$. The dimensional time scale is given by
\begin{equation}
    t^*=\frac{\varepsilon^2R}{V}T=\frac{\mu_g^{2/3}R^{1/3}}{\rho_l^{2/3}V^{5/3}}T.
\end{equation}
For a 1 mm water droplet in air with impact velocity 1 m s$^{-1}$, the touchdown delay incurred here is 11.4 $\mu$s.

The slower closing of the air gap also influences the build-up of pressure beneath the droplet. In figure \ref{fig:FGPfullevve} as the parameters $\tilde{e}_3$ and $\tilde{e}_5$ are decreased in magnitude, the bifurcating behaviour of the pressure profiles is still present. However, the pressure peak amplitude as touchdown is approached is lowered by the introduction of surface deformability, with larger surface deformations resulting in lower pressures. 
\begin{figure*}
    \centering
    \includegraphics[width=15cm]{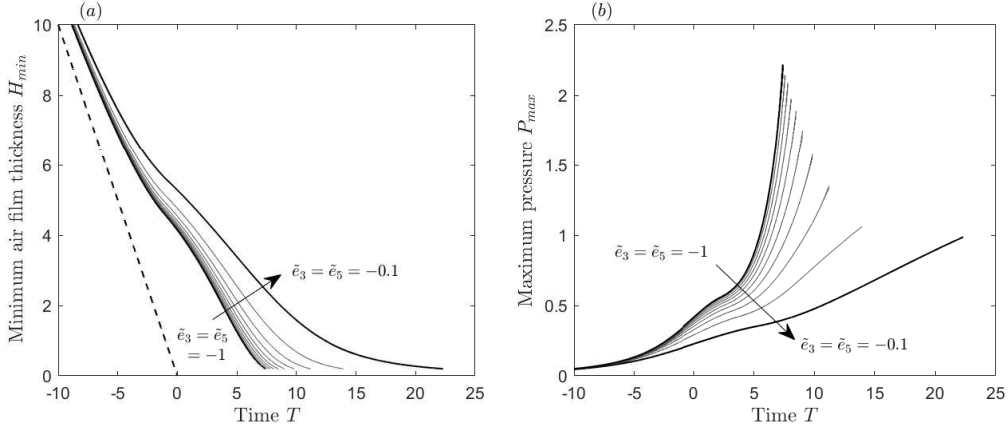}
    \caption{The (\textit{a}) minimum air film thickness and the (\textit{b}) maximum pressure as a function of time $T$ for viscoelastic surfaces with coefficients $\tilde{e}_3=\tilde{e}_5$ ranging from -1 to -0.1 in intervals of 0.1. The dashed line in figure (\textit{a}) corresponds to the minimum film thickness in the absence of air cushioning.}
    \label{fig:minfilmmaxp}
\end{figure*}

Figure \ref{fig:minfilmmaxp} highlights the link between the rate at which the air gap closes and the maximum pressure. The lower pressures seen for more deformable surfaces are found to be present at all time. 
\begin{figure} 
    \centering
    \includegraphics[width=11cm] {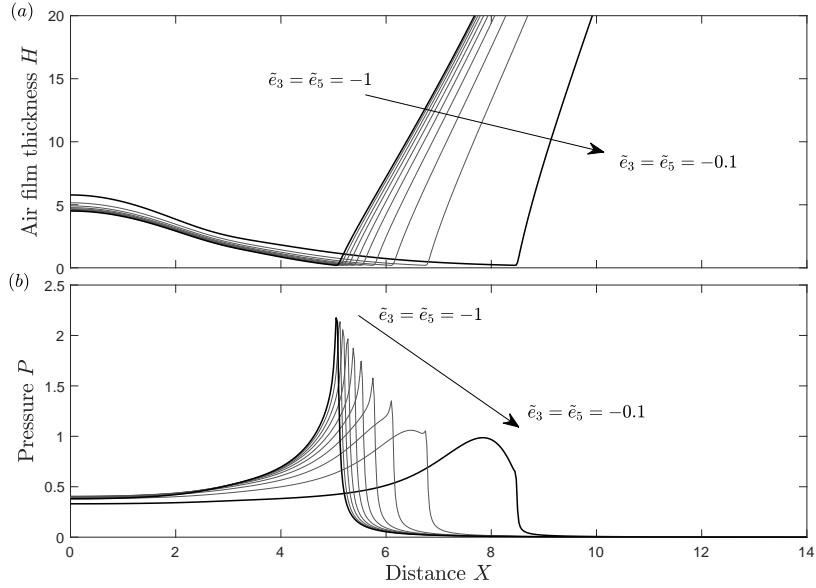}
    \caption{As figure \ref{fig:minfilmmaxp}, except showing (\textit{a}) the air film thickness $H=F-G$ and (\textit{b}) the pressure near touchdown.}
    \label{fig:HP_tdcomp}
\end{figure}

It is also interesting to note the shape of the pressure solution as touchdown is approached. Figure \ref{fig:HP_tdcomp} shows the air film thickness $H=F-G$ and pressure profile solution for a range of values $\tilde{e}_3=\tilde{e}_5$. It can be seen that for surface parameters of sufficiently high magnitude, the pressure profile solution near touchdown is virtually identical to that of the flat surface solution \citep{smith2003,hicks2017}, with a very sharp pressure peak located at the cusp of the air film thickness $H$. As the magnitude of the surface parameters is decreased the sharp pressure peak near touchdown still exists up until around $\tilde{e}_3=\tilde{e}_5=-0.2$ (this solution is also shown in figure \ref{fig:FGPfullevve}) where a rounder pressure peak located just behind the cusp of the air film thickness $H$ is seen. As the magnitude of the surface parameters is decreased further to $\tilde{e}_3=\tilde{e}_5=-0.1$, the rounder pressure peak overtakes the sharp one near touchdown. The location of the lower, rounder pressure peak just behind the cusp of the air film thickness, as opposed to at it, could help explain the extended air gliding behaviour for droplet impacts on to softer solids seen in experiments by \citet{langley2020}. Air gliding is when a droplet skids on a thin air film as opposed to touching down; the numerical solutions of the air film thickness in figure \ref{fig:HP_tdcomp}(\textit{a}) appear to show the onset of such behaviour.

\subsection{Entrapped air size} \label{air}

A vitally important outcome of the dynamics described here is the area of entrapped air underneath the droplet at impact. Trapped air at impact can cause potential problems when it comes to using droplets to spray coat materials or in cooling processes. From the numerical results of pre-impact air-water-substrate interaction in figure \ref{fig:FGPfullevve} it is clear that air entrapment still occurs. What is perhaps more clear in figure \ref{fig:HP_tdcomp} is that the presence of substrate deformability in the viscoelastic model leads to an increase in the area of air entrapped at impact. The framework of our model is in two dimensions, and it is realised that droplet impact is clearly a three dimensional phenomenon. Despite this, we are able to use our model to make a qualitative assessment of the size of entrapped air for variations in the substrate stiffness and (viscous) damping.

The X-wise symmetry of the solutions allows us to calculate the dimensionless area of entrapped air $B$ as
\begin{equation} \label{eq:bint}
    B = 2\int_0^{X_d}H(X,T_d)\mathrm{d}X,
\end{equation}
where $X_d$ is the positive $X$ station of minimum air film thickness and $T_d$ is the touchdown time. It should be stressed that, numerically, touchdown where $H=0$ is never quite realised due to the parabolic degeneracy of the lubrication equation (\ref{eq:lube}). Therefore, the time of touchdown has to be pre-determined as a point when the air film thickness reduces to a very small positive value. Numerically, the smaller the grid size and time step in the numerical scheme, the smaller we can make this value. However, this has to be balanced with the increased computational cost due to the huge amount of simulations needed to be run to build a parametric picture. In the present section we therefore set this pre-determined value of air film thickness, where the droplet is considered to have reached touchdown, as $H_{min}=0.1$ (the solutions in figures \ref{fig:FGPfullevve} and \ref{fig:HP_tdcomp} are plotted up until $H_{min}=0.2$, for comparison).

\begin{figure}
    \centering
    \includegraphics[width=11cm]{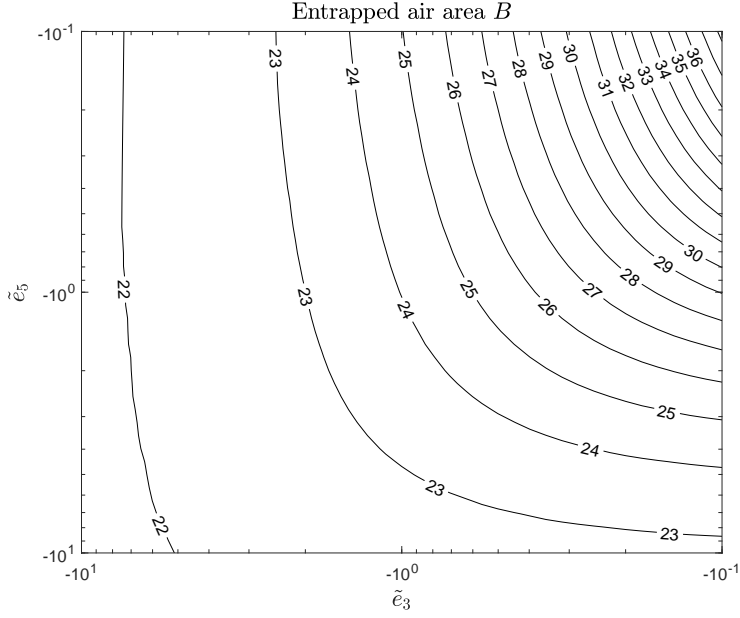}
    \caption[width=\textwidth]{Parametric plot of the entrapped air area $B$ for variations in the parameters $\tilde{e}_3$ and $\tilde{e}_5$ in the viscoelastic model.}
    \label{fig:bce3e5}
\end{figure}

Figure \ref{fig:bce3e5} shows how the dimensionless area of entrapped air $B$ depends on the surface parameters $\tilde{e}_3$ and $\tilde{e}_5$, for the viscoelastic surface model. Clearly, for any decrease in magnitude of $\tilde{e}_3$ or $\tilde{e}_5$ there is an increase in the area of entrapped air. The dimensional entrapped air area $b^*$ can be related to the dimensional area $B$ by applying the vertical and horizontal length scales, which yields
\begin{equation} \label{eq:bdim}
    b^* = \varepsilon^3R^2B = \frac{\mu_gR}{\rho_lV}B,
\end{equation}
where the pre-factor $B$ is, in the viscoelastic model, a function of the parameters $\tilde{e}_3$ and $\tilde{e}_5$ and its dependency is shown in figure \ref{fig:bce3e5} for parameters in the range [-10,-0.1]. For parameter values of magnitude lower than 0.1, the area of entrapped air continues to increase and for parameters of magnitude higher than 10 the behaviour of the droplet is identical to that of a flat rigid surface.

Equation (\ref{eq:bdim}) for the dimensional entrapped air area highlights the importance of the droplet velocity and radius on the size of entrapped air at impact. For a flat rigid surface, where $B$ is a given constant, droplets with a larger radius will entrap more air due to having a larger free-surface to interact with the air prior to impact and droplets with a larger velocity will entrap less air due to the droplet having less time to deform prior to impact. In the viscoelastic model, the numerical pre-factor $B$ depends on $\tilde{e}_3$ and $\tilde{e}_5$ and formulas for these parameters are given in (\ref{eq:parameters}). It is interesting to note that the parameter $\tilde{e}_3$ does not depend on the droplet radius $R$. It is therefore of interest to see how the size of entrapped air varies with droplet velocity and radius for given fluid and structural parameters (the structural parameters of concern here are the stiffness $\kappa^*$ and the damping $C^*$).

\begin{figure}
    \centering
    \includegraphics[width=11cm]{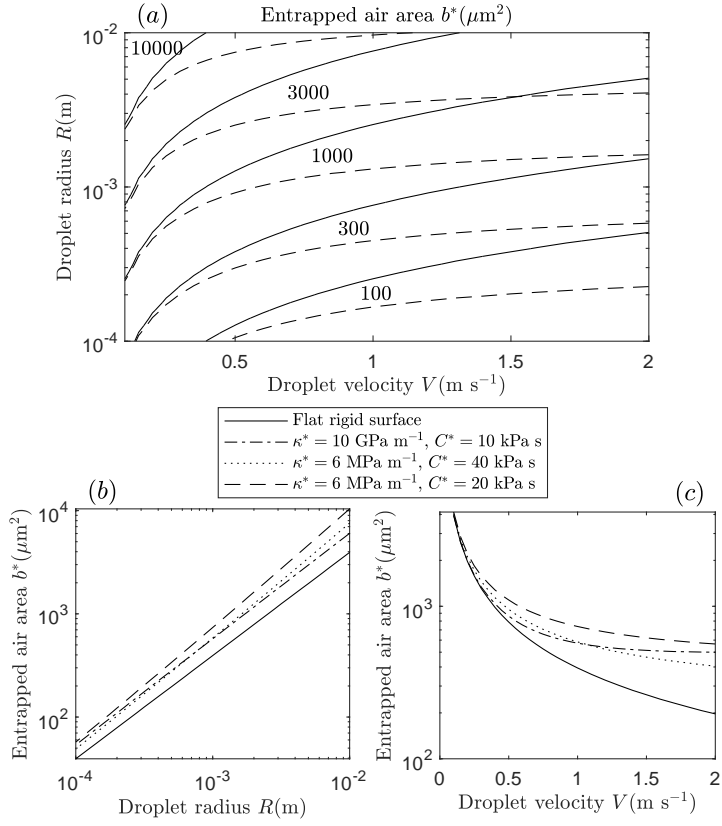}
    \caption[width=\textwidth]{For impact of a water droplet in air, (\textit{a}) variations in the entrapped air area $b^*$($\mu$m$^2$) as the droplet velocity $V$(m s$^{-1}$) and radius $R$(m) vary for a flat rigid surface (solid line) and a soft viscoelastic surface with spring stiffness $\kappa^*=6$ MPa m$^{-1}$ and (viscous) damping $C^*=20$ kPa s (dashed line), (\textit{b}) entrapped air area for a droplet velocity of 1 m s$^{-1}$ and variations in droplet radius and (\textit{c}) entrapped air area for a droplet of radius 1 mm and variations in droplet velocity. In figures (\textit{b}-\textit{c}) results are presented for a variety of spring stiffness $\kappa^*$ and (viscous) damping $C^*$. }
    \label{fig:RVbubcontours}
\end{figure}

Figure \ref{fig:RVbubcontours} gives a comparison produced by results for the dimensional entrapped air area of a water droplet in air over wide ranges of the droplet radius $R$ and velocity $V$, for a flat rigid surface and soft viscoelastic surfaces. We chose to vary the droplet velocity from 0.1 m s$^{-1}$ to 2 m s$^{-1}$ and the radius from $10^{-4}$ m to $10^{-2}$ m in order to keep the parameter regime mostly contained in the valid model parameter regime given in figure \ref{fig:parareg}. Figure \ref{fig:RVbubcontours}(\textit{a}) shows the variations of the entrapped air area $b^*$ as the droplet velocity and radius vary, comparing a flat rigid surface to a soft viscoelastic surface with spring stiffness $\kappa^*=6$ MPa m$^{-1}$ and (viscous) damping $C^*=20$ kPa s. For all combinations of droplet velocity and droplet radius, more air is entrapped by the soft viscoelastic surface, as is to be expected. For small droplet impact velocities, the effect of the soft solid is minimal and it behaves much like a rigid surface. As the droplet velocity increases, the effect of the soft viscoelastic surface is far more profound. From equation (\ref{eq:bdim}) we can see that for a flat rigid surface the area of entrapped air decreases with increased droplet velocity, whereas for the soft viscoelastic surface this decrease in entrapped air area is delayed and even halted for increased droplet velocity, for the parameters under consideration here. This can be seen more clearly in figure \ref{fig:RVbubcontours}(\textit{c}) where the entrapped air area is given as a function of droplet velocity for a 1 mm water droplet in air, for a variety of spring stiffness $\kappa^*$ and (viscous) damping $C^*$. This is due to higher air film pressures, induced by higher droplet velocities, causing larger surface deformation prior to impact. By considering the shape of the contour lines in figure \ref{fig:RVbubcontours}(\textit{a}), variations of the droplet radius have a less profound influence on the increase in entrapped air on a soft viscoelastic solid compared to a flat rigid surface. This is shown more clearly in figure \ref{fig:RVbubcontours}(\textit{b}), where the entrapped air area is plotted as a function of droplet radius for an impact velocity of 1 m s$^{-1}$. On a logarithmic scale, the increased amount of entrapped air due to impact onto a soft viscoelastic surface is almost independent of droplet radius.

\begin{figure*}
    \centering
    \includegraphics[width=15cm]{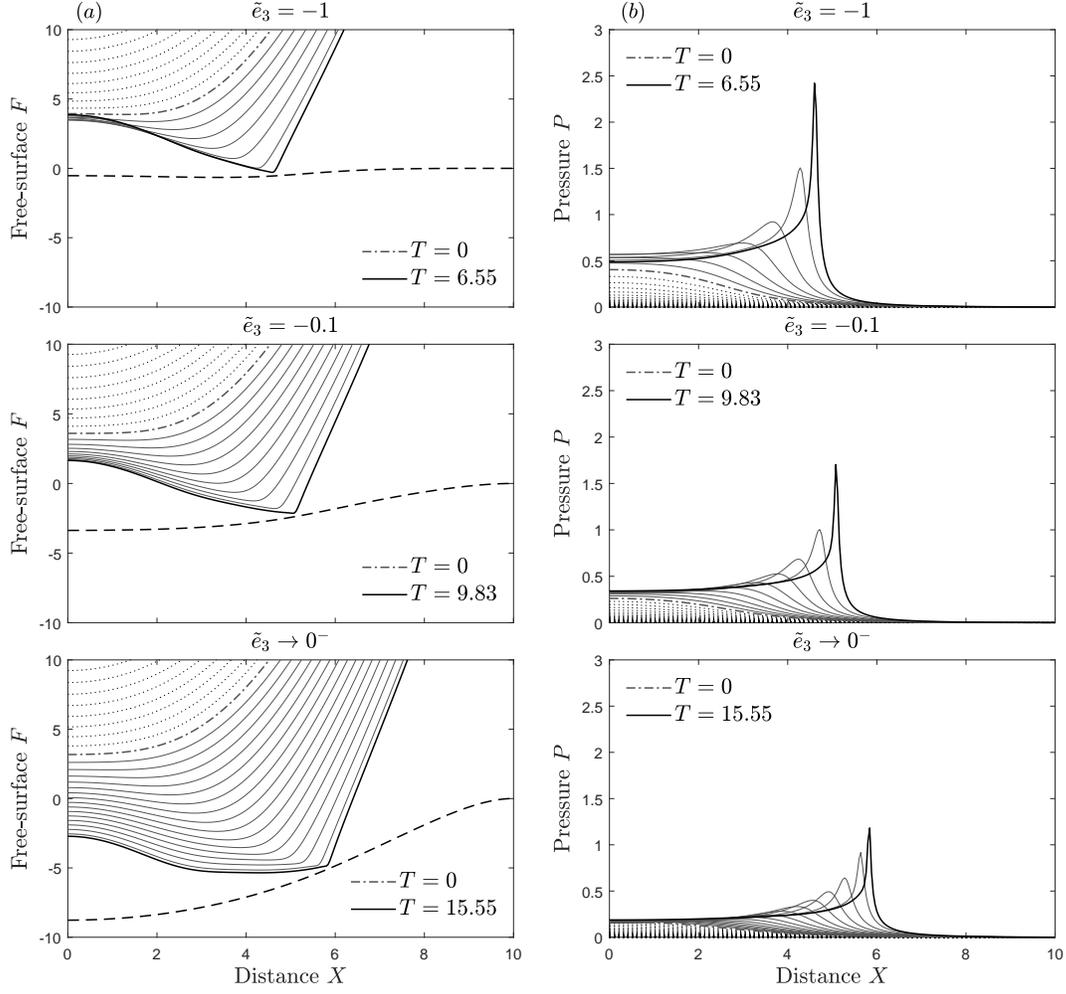}
    \caption{Solution profiles, showing evolution of (\textit{a}) the free-surface height $F$ and (\textit{b}) the pressure $P$ for normal impact of a droplet onto a flexible surface with parameters $(\tilde{e}_1,\tilde{e}_2,\tilde{e}_3,\tilde{e}_4,\tilde{e}_5)=(-1,1,\tilde{e}_3,-1,0)$ and $\tilde{e}_3=-1$, $\tilde{e}_3=-0.1$ and $\tilde{e}_3\rightarrow 0^-$. The boundary of the flexible surface is at $[X_1,X_2]=[-10,10]$. The  solutions  are  shown  in integer time increments except for the final thick solid line, which is the solution just prior to touchdown, and the dashed line, which is the flexible surface solution $G$ just prior to touchdown. The dash-dotted line is the solution at $T=0$ where the droplet would touchdown on an undisturbed surface in the absence of air cushioning.}
    \label{fig:FGPfullevflex}
\end{figure*}

\begin{figure*}
    \centering
    \includegraphics[width=15cm]{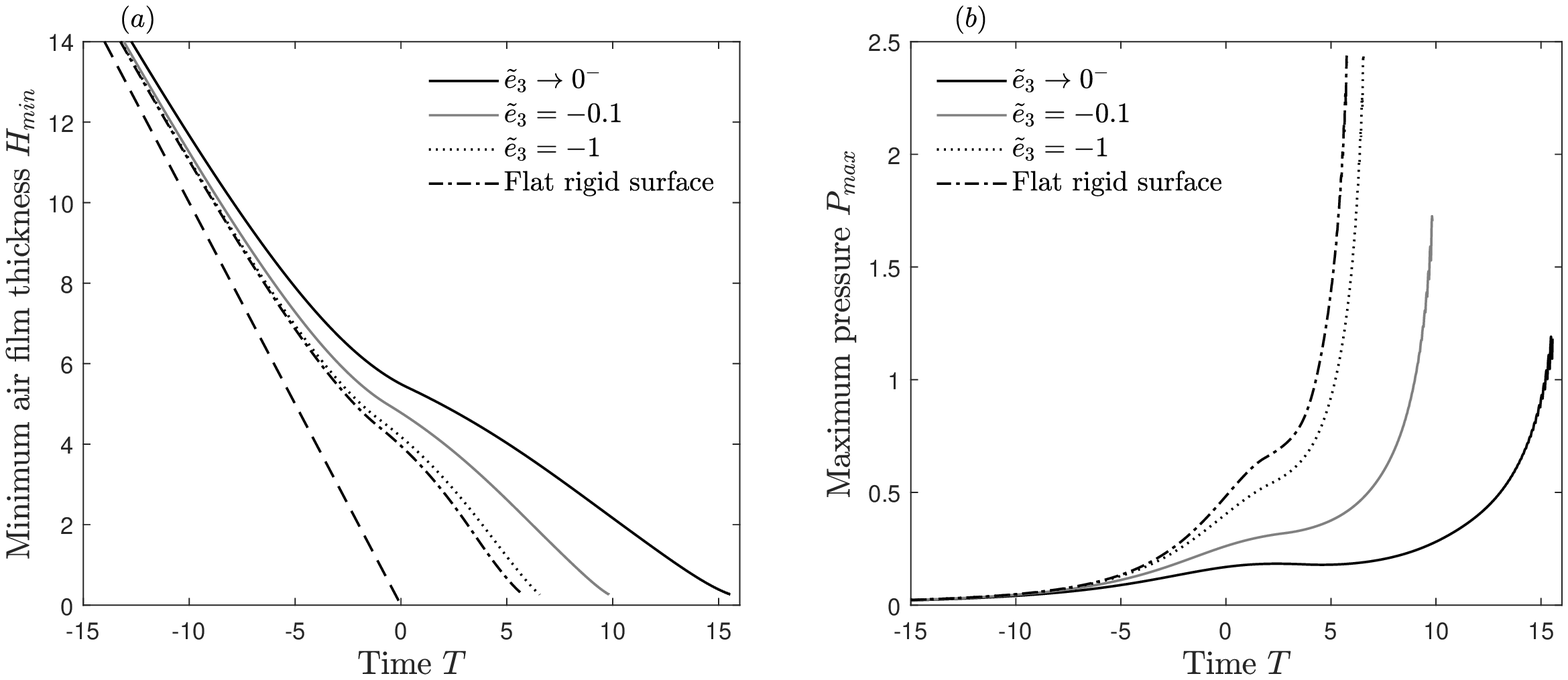}
    \caption{The (\textit{a}) minimum air film thickness and the (\textit{b}) maximum pressure as a function of time $T$ for a flat rigid surface and flexible surfaces with coefficients $(\tilde{e}_1,\tilde{e}_2,\tilde{e}_3,\tilde{e}_4,\tilde{e}_5)=(-1,1,\tilde{e}_3,-1,0)$, a range of values of $\tilde{e}_3$ and boundary $[X_1,X_2]=[-10,10]$. In (\textit{a}), the dashed line corresponds to the solution in the absence of air cushioning.}
    \label{fig:HminPmax_flex}
\end{figure*}

\section{Impact onto a flexible substrate}\label{flexible}

We now turn our attention to analysing the air cushioning effect in droplet impact onto a more general flexible surface, where all terms except for the damping term in equation (\ref{eq:flex}) are retained. In practise, the magnitude of the dimensionless parameters $\tilde{e}_i$ can vary dramatically and over a number of orders of magnitude. By considering figure $\ref{fig:parareg}$ we can see the range of droplet velocities and radii where our model is applicable, and it is the structural parameters that will require special attention. Explicit formulae for the parameters $\tilde{e}_i$ are given in (\ref{eq:parameters}) and if we consider a water droplet of radius 1 mm and impact velocity 1 m s$^{-1}$ we are able to obtain an order of magnitude estimate of the dimensionless parameters $\tilde{e}_i$ in terms of their corresponding structural coefficient,
\begin{equation}
    (|\tilde{e}_1|,|\tilde{e}_2|,|\tilde{e}_3|,|\tilde{e}_4|)\sim(10^7B^*,10^{-2}T_t^*,10^{-11}\kappa^*,10^1M^*).
\end{equation}
Therefore, if we wish for the dimensionless surface parameters $\tilde{e}_i$ to be of order one, plus or minus a few orders of magnitude, then we now know what order of magnitude the dimensional structural parameters need to be in our model. If we wish to consider the rather large range of parameter magnitudes $10^{-4}<|\tilde{e}_i|<10^4$, then this implies our structural parameters are of the order $10^{-11}<B^*<10^{-3}$, $10^{-2}<T_t^*<10^6$, $10^7<\kappa^*<10^{15}$ and $10^{-5}<M^*<10^{3}$. Rather unsurprisingly, we require a flexural rigidity $B^*$ of small magnitude. The flexural rigidity can be defined by the formula
\begin{equation}
    B^*=\frac{Eh^3}{12(1-\nu^2)},
\end{equation}
where $E$ is the Young's Modulus of the material, $h$ is the elastic thickness and $\nu$ is the Poisson ratio. Therefore, for the flexural rigidity $B^*$ to be of a suitably small magnitude, we would require a combination of a relatively low Young's Modulus combined with thin plates. Examples of impacts involving such a combination exist in nature, such as impact onto leaves \citep{gilet2015}, and for impact onto thin foils of certain materials \citep{contino2019}. 

The surface shape equation (\ref{eq:flex}) is then solved numerically, again coupled with equation (\ref{eq:cauchy}) and (\ref{eq:lube}). The boundary conditions and the numerical treatment are slightly different from \ref{visco}, due to the spatial derivatives now present in the surface equation. Because of the spatial derivatives, boundary conditions for the surface shape $G$ are required now at fixed positions. We choose clamped plate boundary conditions, which are
\begin{equation}
    G=\frac{\partial G}{\partial X}=0,~~~~~\mbox{at\ } G=X_1,X_2,
\end{equation}
where $X_1,X_2$ are given $X$ stations and will be stated in each case. Typically we take $X_1,X_2$ as the end points of the computational domain. The same efficient numerical scheme applied in \ref{visco} cannot practically be used here because of the increased numerical complexity due to sixth order derivatives occurring when substituting the surface equation (\ref{eq:flex}) into the lubrication equation (\ref{eq:lube}). We therefore adopt a more standard approach of iterating between each equation (\ref{eq:cauchy}), (\ref{eq:lube}) and (\ref{eq:flex}) and solving for $P$, $F$ and $G$, respectively, until convergence, using the same algorithms outlined in \ref{visco}.

\subsection{Free-surface and pressure profile solution}

Figure \ref{fig:FGPfullevflex} shows the solution profiles for the free-surface $F$ and the pressure $P$ for three different values of the stiffness parameter $\tilde{e}_3$, with the other surface parameters fixed. In practice zero stiffness is not possible, hence, while all the other parameters are fixed, $\tilde{e}_3\rightarrow0^-$ is a limiting scenario (the same applies for the other parameters in \ref{airflex}). The dashed line in each plot of figure \ref{fig:FGPfullevflex}(\textit{a}) corresponds to the solution of the flexible surface shape $G$ as touchdown is approached. The results illustrate the outcomes of variations in the stiffness parameter $\tilde{e}_3$, however variations in any of the other surface parameters yield qualitatively similar results. The solution in the flat surface case is not shown here for brevity, see figure \ref{fig:FGPfullevve} for comparisons.

\begin{figure}
    \centering
    \includegraphics[width=11cm]{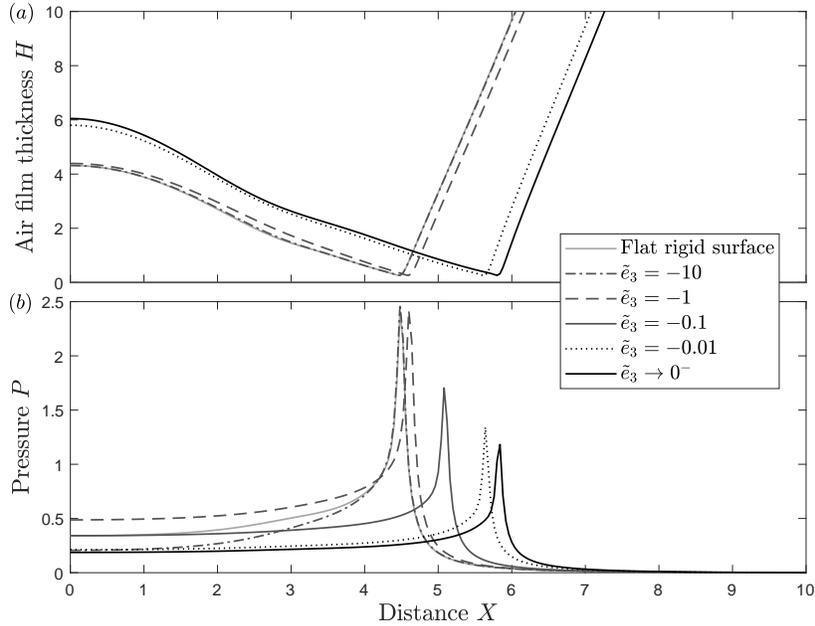}
    \caption{As figure \ref{fig:HminPmax_flex}, except showing (\textit{a}) air film thickness $H$ and (\textit{b}) pressure $P$ near touchdown.}
    \label{fig:HP_td_flex}
\end{figure}

The aim of this section is to highlight the effect surface flexibility can have on the pre-impact phase of droplet impact, as opposed to the effect of a soft deformable surface considered in \ref{visco}. Figure \ref{fig:FGPfullevflex} shows that much of the same conclusions can be drawn. First, increased surface flexibility leads to a delay to touchdown. This delay to touchdown acts to further decelerate the droplet free surface and results in a lower pressure build-up underneath the droplet. The pressure peak amplitude is again lower near touchdown for more flexible surfaces. Figure \ref{fig:HminPmax_flex} shows the minimum air film thickness and the maximum pressure as functions of time, where it can be seen that this lower pressure buildup is present for all time. Figure \ref{fig:HP_td_flex} compares the solutions of the air film thickness and the pressure near touchdown, showing clearly the reduced pressure peak and increased air entrapment near touchdown for reductions in magnitude of the stiffness parameter $\tilde{e}_3$.

The main difference to highlight when comparing air cushioning behaviour of droplet impact on a flexible surface to impact on a soft viscoelastic surface is the shape of the pressure profile upon touchdown. Here, in figure \ref{fig:FGPfullevflex}, reductions in the magnitude of the parameter $\tilde{e}_3$ lead to reductions in the amplitude of the pressure peak near touchdown, but the peak remains sharp. This is unlike the finding for the viscoelastic surface, where increased deformability leads to a decrease in the pressure peak amplitude and also a rounder, softer peak. This could be due, in part, to the more abrupt approach to touchdown seen in figure \ref{fig:HminPmax_flex}, for a flexible surface, compared with the gentler approach seen in figure \ref{fig:minfilmmaxp}, for a soft viscoelastic surface. Also, the droplet free surface cusp remains relatively sharp in the flexible surface case as deformability increases, while in the viscoelastic case it becomes rounded, which is likely to play a key role in the shape of the pressure peak. However, as the droplet free-surface becomes sharp, surface tension effects will become significant and would need to be considered here in the flexible surface case. At leading order, the local touchdown behaviour here is identical to that described in \citet{smith2003}.

\subsection{Entrapped air size} \label{airflex}

Using equation (\ref{eq:bint}) we are again able to make a qualitative assessment of how variations in the surface properties affect the size of entrapped air. Here, we perform a study of how individual variations in the parameters $\tilde{e}_1$, $\tilde{e}_2$, $\tilde{e}_3$ and $\tilde{e}_4$ (with $\tilde{e}_5=0$) alter the area of entrapped air at touchdown. The size of the air gap at which the calculation of equation (\ref{eq:bint}) is performed is 0.26 for this model, which is the smallest achievable for all compared results and larger than that considered in \ref{visco} due to the more difficult numerical task. 

\begin{figure}
    \centering
    \includegraphics[width=11cm]{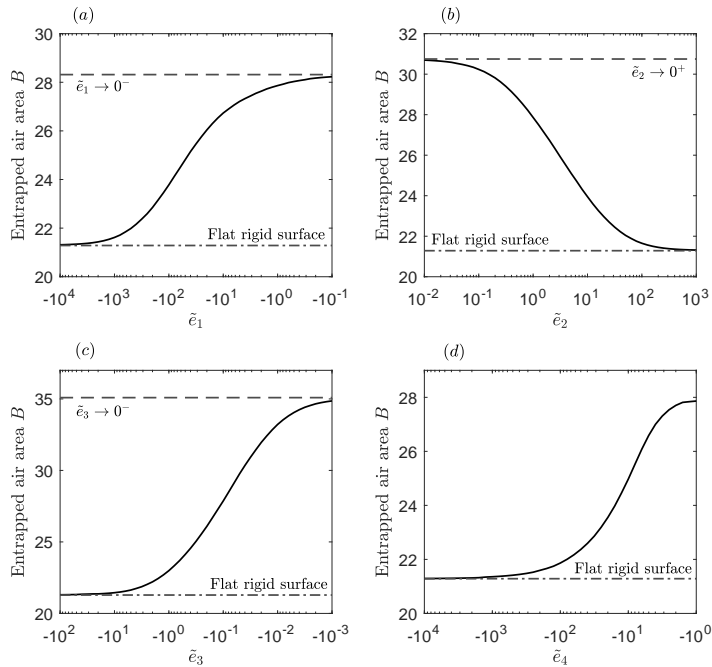}
    \caption{Area of entrapped air $B$ for droplet impact onto flexible surfaces, with a boundary of $[X_1,X_2]=[-10,10]$ and surface parameters (\textit{a}) $(\tilde{e}_1,\tilde{e}_2,\tilde{e}_3,\tilde{e}_4,\tilde{e}_5)=(\tilde{e}_1,1,-0.1,-1,0)$ for variations in $\tilde{e}_1$, (\textit{b}) $(-1,\tilde{e}_2,-0.1,-1,0)$ for variations in $\tilde{e}_2$, (\textit{c}) $(-1,1,\tilde{e}_3,-1,0)$ for variations in $\tilde{e}_3$ and (\textit{d}) $(-1,1,-0.1,\tilde{e}_4,0)$ for variations in $\tilde{e}_4$. In each figure, the dash-dotted line corresponds to the value for a flat rigid surface and the dashed line corresponds to the limiting value as the relevant parameter tends to zero. In (\textit{d}) the limit as the mass density parameter $\tilde{e}_4$ tends to zero cannot be found, due to an unbounded surface velocity.}
    \label{fig:e1e2e3e4bub}
\end{figure}

Figure \ref{fig:e1e2e3e4bub} shows how individual variations in the surface parameters $\tilde{e}_1$, $\tilde{e}_2$, $\tilde{e}_3$ and $\tilde{e}_4$ affect the entrapped air area at touchdown. Logarithmic variations of each parameter lead to a similar dependency on the entrapped air area. Each figure exhibits plateauing behaviour to the flat rigid surface value for large magnitude parameters and to the value for the solution as the parameter tends to zero, except for figure \ref{fig:e1e2e3e4bub}(\textit{d}), where, in the current setting, variations of the magnitude of the mass density parameter $\tilde{e}_4$ much lower than 1 lead to an unbounded surface velocity $\partial G/\partial T$ at some point. From figure \ref{fig:e1e2e3e4bub} it is clear that any increase in flexibility of the substrate leads to an increase in entrapped air.

\section{Connection with experiments} \label{expcon}

In \ref{visco} and \ref{flexible} we performed a numerical study into the air cushioning phase of droplet impact onto deformable surfaces. Here, we now attempt to make connections to recent experiments. Due to our work being two dimensional and approximate, using a number of assumptions on the fluid and structural parameters, these connections are tentative, but they seem encouraging nonetheless.

\subsection{Entrapped air}

In \citet{langley2020} it was found, experimentally, that droplet impacts onto softer, more deformable, solids would entrap more air. They performed experiments using ultra-high-speed interferometry to capture the droplet free surface at impact, and varied the droplet velocity and the surface stiffness. We have qualitative agreement with these findings in our analytical results. We found in \ref{visco} and \ref{flexible} that reductions in surface stiffness resulted in an increase in entrapped air. A more subtle point to make from our analytical work is the non-intuitive dependency of the size of entrapped air on the droplet velocity. On a rigid flat surface, the size of entrapped air will decrease for higher velocities, whereas for impact onto a soft viscoelastic surface this reduction is delayed and even halted for increased impact velocities. This is alluded to in \citet{langley2020} also and, although described in terms of gas compression while our work is incompressible, they show that entrapped air can increase with increased droplet velocity for impact onto soft solids.

\subsection{Implications on splashing}

\citet{howland2016} focused their study on how soft deformable surfaces affect droplet splashing. They performed experiments by impacting ethanol droplets on silicone or acrylic substrates of varying stiffnesses, and found that the lower the stiffness, the less likely the droplet was to splash. This was found to be most likely due to higher stiffness substrate having sheet ejection of higher velocity, resulting in the sheet leaving the surface and breaking up, forming a corona splash. By contrast, for less stiff surfaces, the ejected sheet is of lower velocity and does initially leave the substrate, but can fall back down onto the substrate and slow down, suppressing the splash. They were unable to investigate the sheet ejection experimentally, because of the small time and length scales associated with it. Hence, they investigated it using numerical simulation, and found that lowering the surface stiffness reduced the contact pressure just before the sheet is ejected. Our results in section \ref{visco} and \ref{flexible} show clearly that reducing the surface stiffness leads to a reduction in the pressure peak just prior to impact. In particular, the results in \ref{visco} for droplet impact onto a viscoelastic solid show a softening of the pressure peak prior to touchdown. Although our study is solely focused on pre-impact behaviour, this reduction in pressure appears important in understanding the reduced sheet ejection speed mentioned in \citet{howland2016}, which results in potential splash suppression.

\section{Large surface deformation} \label{largeflex}

In this section we examine the effect of large surface deformations on the system. Here, we will consider the system in the limit $\tilde{e}_1\rightarrow0$, $\tilde{e}_2\rightarrow0$, $\tilde{e}_3\rightarrow0$ and $\tilde{e}_5\rightarrow0$, where equation (\ref{eq:flex}) may be written simply as $\tilde{e}_4\partial^2G/\partial T^2=P$. What this allows us to do is reduce our previous system of three equations to two, by writing $H=F-G$, the air film thickness. The new system is now
\begin{subequations} \label{eq:HP}
\begin{equation}
\frac{\partial^2 H}{\partial T^2} = \frac{1}{\pi}\dashint_{-\infty}^{\infty}\frac{\partial P}{\partial \zeta}(\zeta,T)\frac{\mathrm{d}\zeta}{X-\zeta}-\frac{P}{\tilde{e}_4}, \\
\end{equation}
\begin{equation}
\frac{\partial}{\partial X}\left(H^3\frac{\partial P}{\partial X}\right) =12\frac{\partial H}{\partial T}.
\end{equation}
\end{subequations}

Equations (\ref{eq:HP}) can be further simplified by rescaling all the variables to account for $|\tilde{e}_4|$. We are interested in the case of $\tilde{e}_4$ being small (and negative) and the time scale being large, as for large surface deformations we expect touchdown to be further delayed. Hence we take the following rescaling,
\begin{equation}
    (H,P,X,T) = (|\tilde{e}_4|^{-1}\bar{H},|\tilde{e}_4|^2\bar{P},|\tilde{e}_4|^{-1/2}\bar{X},|\tilde{e}_4|^{-1}\bar{T});
\end{equation}
then equations (\ref{eq:HP}) reduce to, dropping the overbar notation,
\begin{subequations} \label{eq:HPsc}
\begin{equation} \label{eq:HPsc_ps}
\frac{\partial^2{H}}{\partial{T}^2}={P}, 
\end{equation}
\begin{equation}
\frac{\partial}{\partial {X}}\left({H}^3\frac{\partial{P}}{\partial {X}}\right)=12\frac{\partial {H}}{\partial {T}},
\end{equation}
\end{subequations}
where the relative error in ignoring the Cauchy integral is $O(|\tilde{e}_4|^{3/2})$, which is suitably small. We impose
\refstepcounter{equation} 
$$
    {H}\sim\frac{{X}^2}{2}-{T},~~~~~{P}\sim0,~~~~~\mbox{as\ }|{X}|\rightarrow\infty~\mbox{or\ }{T}\rightarrow -\infty
    \eqno{(\theequation{\text{a},\text{b}})} \label{eq:HbPbff}
$$
as the far-field conditions.

As was first remarked in \citet{korobkin2008} for a different model, the pressure-shape law (\ref{eq:HPsc_ps}) can be viewed as an alternative to the Cauchy-Hilbert law in \citet{smith2003}, such as in pressure-displacement laws in interacting boundary layers \citep{smith1982,sobey2000}. The new coupled system (\ref{eq:HPsc}) allows us to make far more analytical progress than for the previous system examined computationally in \ref{visco} and \ref{flexible}.

\subsection{Computational solutions}

The coupled system (\ref{eq:HPsc}) was solved numerically using a scheme identical to that outlined in \ref{visco}, with the new local pressure-shape relationship (\ref{eq:HPsc_ps}). As is to be expected, this scheme is far faster than the one involving the Cauchy-Hilbert integral. An appropriate spatial domain here was found to be $[-20,20]$.

\begin{figure*}
    \centering
    \includegraphics[width=15cm]{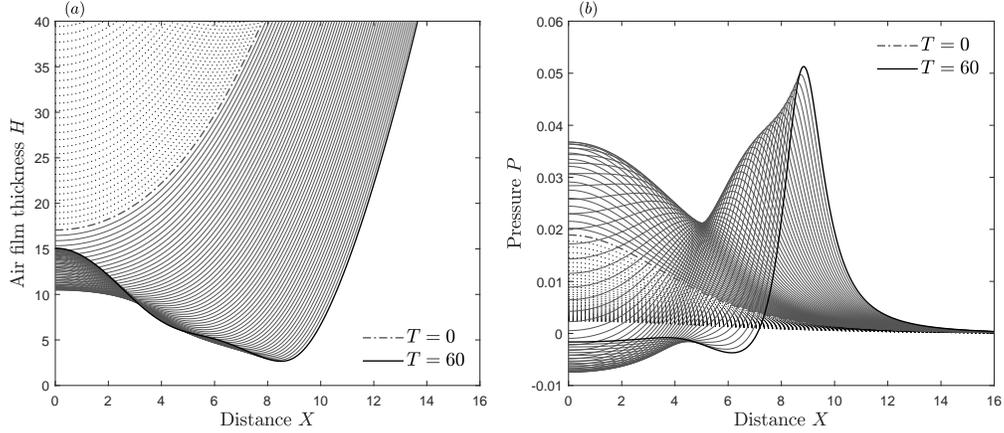}
    \caption{Solution profiles of the ($\mathit{a}$) air film thickness $H$ and ($\mathit{b}$) pressure $P$. The solutions are shown in integer time increments up until $T=60$.}
     \label{fig:HbarPbar}
\end{figure*}

Figure \ref{fig:HbarPbar} shows the time-marched solutions of ${H}$ and ${P}$. The mechanisms are essentially the same as before. The droplet is released at some suitably large negative time and begins to approach the surface. The air film thickness is initially parabolic and as the air film thickness begins to decrease, the pressure begins to rise in a single peak and results in the air film thickness deviating from the parabolic shape. The pressure then starts to have two peaks as the air film thickness approaches zero in two different locations equidistant from the center line. 

The solutions of ${H}$ and ${P}$ in Figure \ref{fig:HbarPbar} show clearly two traits of droplet impact on a very deformable surface that can be intuitively expected from the results in \ref{visco} and \ref{flexible}. Figure \ref{fig:HbarPbar} shows solutions up until $T=60$. The longer time scales associated here are clear to see and it is expected that touchdown is not reached in finite time. We can also see that the horizontal bubble extent has significantly increased.

\subsection{Large time behaviour}

The computational results given in figure \ref{fig:HbarPbar} show that the time scales involved in this system are large and suggest that perhaps touchdown in finite time is not reached on this time scale. Hence, we seek a solution at large positive time.  Suppose that the scaled air thickness and pressure are rising relatively fast in spatial terms, near a region where $X=cT^\alpha$ say, with $\alpha$ a positive constant. The length scaling in this region then takes the form
\begin{equation} \label{eq:xi}
    X = cT^\alpha+T^m\xi,~~~~~\mbox{as\ } T\rightarrow\infty,
\end{equation}
where $\xi$ is order unity, the constants $\alpha$ and $m$ are unknown and in order for the region to be local we require that $m< \alpha$. Expansions of $H$ and $P$ are then taken in the rather general form
\begin{subequations} \label{eq:HPexp}
\begin{equation} \label{eq:HPexpH}
H \sim T^\lambda\hat{H}(\xi),
\end{equation}
\begin{equation}
P \sim T^{\lambda+2\alpha-2m-2}\hat{P}(\xi)
\end{equation}
\end{subequations}
where $\lambda$ is an unknown constant. It is to be expected that $\lambda\leq0$ as $H$ is not growing in time, and the expansion of $P$ is inferred from seeking a balance in equation (\ref{eq:HPsc_ps}). Substitution of the expansions (\ref{eq:HPexp}) into the governing equations (\ref{eq:HPsc}) leads to the system
\begin{subequations} \label{eq:HPhatsyt}
\begin{equation} \label{eq:HPhatsyt_ps}
\alpha^2c^2 \frac{\mathrm{d}^2\hat{H}}{\mathrm{d}\xi^2} = \hat{P},
\end{equation}
\begin{equation}
\frac{\mathrm{d}}{\mathrm{d}\xi}\left(\hat{H}^3\frac{\mathrm{d}\hat{P}}{\mathrm{d}\xi}\right) = -12\alpha c\frac{\mathrm{d}\hat{H}}{\mathrm{d}\xi},
\end{equation}
\end{subequations}
to leading order, subject to 
\begin{equation}
    m = \lambda + \frac{\alpha-1}{3}
\end{equation}
and also $3\lambda<2\alpha+1$. The system (\ref{eq:HPhatsyt}) can be further reduced to an ordinary differential equation in $\hat{H}$ alone,
\begin{equation} \label{eq:Htd}
    \hat{H}^3\frac{\mathrm{d}^3\hat{H}}{\mathrm{d}\xi^3}=-\frac{12}{\alpha c}\hat{H}+k,
\end{equation}
where $k$ is an integration constant and must be positive in order to keep $\hat{H}$ positive. Equations of the form (\ref{eq:Htd}) occur commonly in a number of applied mathematics problems \citep{kalliadasis1994,braun2003,purvis2004} and it is expected that the solution $\hat{H}$ will tend to a non-zero constant $\alpha ck/12$ as $\xi\rightarrow-\infty$. Perturbations to this constant value occur, such that
\begin{equation}
    \hat{H}\sim\frac{\alpha ck}{12}+\Re[\gamma_1\exp(\gamma_2\xi)]~~~~~\mbox{as\ }\xi\rightarrow-\infty,
\end{equation}
where $\Re$ denotes the real part and $\gamma_1$ and $\gamma_2$ are complex constants. Note, if we were to take the perturbation in the form $\gamma_1\exp(\gamma_2\xi)$, where $\gamma_1$ and $\gamma_2$ are real constants, then that would lead to $\gamma_2<0$ and thus an exponentially growing perturbation. The complex constant $\gamma_2$ satisfies $\gamma_2^3=-q$, where $q=(12/\alpha ck^{3/4})^4$, and the complex constant $\gamma_1$ remains arbitrary. There are then three possible solutions for $\gamma_2$, and we wish to choose the ones such that $\Re[\gamma_2]>0$ so that we have a decaying perturbation. There are two such solutions, $\gamma_2= q^{1/3}(1\pm i\sqrt{3})/2$, which leads to the large negative $\xi$ asymptote taking the form
\begin{widetext}
\begin{equation} \label{eq:largenxi}
    \hat{H}\sim\frac{\alpha c{k}}{12}+\exp\left(\frac{1}{2}q^{1/3}\xi\right) 
    \left\{\gamma_{1,1}\cos\left(\frac{\sqrt{3}}{2}q^{1/3}\xi\right)+\gamma_{1,2}\sin\left(\frac{\sqrt{3}}{2}q^{1/3}\xi\right)\right\} 
    ~~~~~\mbox{as\ }\xi\rightarrow-\infty,
\end{equation}
\end{widetext}
where $\gamma_{1,1}$ and $\gamma_{1,2}$ remain arbitrary, still. For large positive $\xi$ it can be readily shown that 
\begin{equation} \label{eq:largepxi}
    \hat{H}\sim\lambda_1\xi(3\ln\xi)^{1/3}~~~~~\mbox{as\ }\xi\rightarrow\infty,
\end{equation}
where $\lambda_1=(12/\alpha c)^{1/3}$. From equation (\ref{eq:HPhatsyt_ps}) we can also show that the corresponding $\hat{P}$ asymptote is 
\begin{equation} \label{eq:largepxip}
    \hat{P}\sim\lambda_2\xi^{-1}(3\ln\xi)^{-2/3}~~~~~\mbox{as\ }\xi\rightarrow\infty,
\end{equation}
where $\lambda_2=\alpha^2 c^2\lambda_1$.

Equation (\ref{eq:Htd}) was then solved numerically as a boundary value problem using Newton-Raphson iterations, with the asymptotes for $\hat{H}$ (\ref{eq:largenxi}-\ref{eq:largepxi}) imposed at suitable large negative and positive $\xi$ values, respectively. The corresponding solution for $\hat{P}$ is then calculated from equation (\ref{eq:HPhatsyt_ps}) using central finite differences. Without loss of generality, the constants $\alpha^2c^2$, $12/(\alpha c)$ and $k$ can be normalised to unity by a division of $\hat{H}$, $\xi$, $\hat{P}$ by $\alpha ck/12$, $(\alpha c/12)^{4/3}k$, $(12^5\alpha c)^{1/3}/k$ in turn.
\begin{figure*}
    \includegraphics[width=15cm]{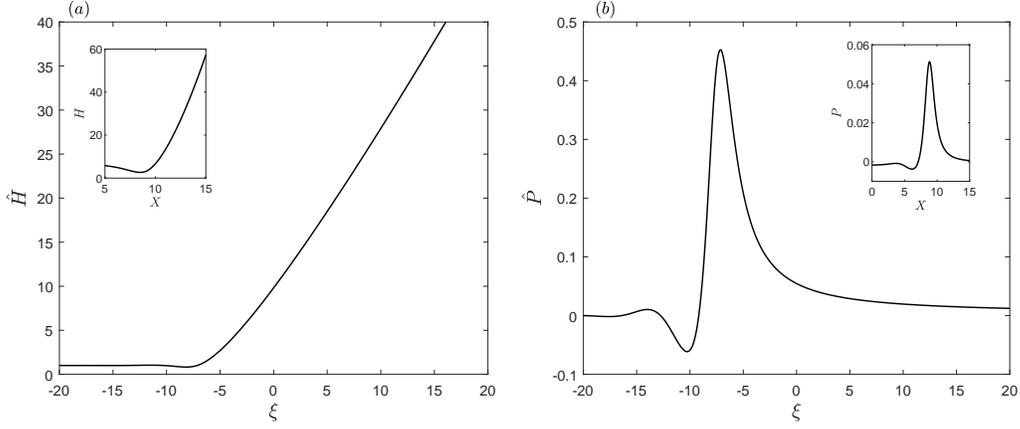}
    \caption{Solution of equation ($\mathit{a}$) (\ref{eq:Htd}) for $\hat{H}$, and then ($\mathit{b}$) (\ref{eq:HPhatsyt}$\mathit{a}$) for $\hat{P}$, given $\hat{H}$, in normalised form for $\gamma_{1,1}=\gamma_{1,2}=1$. The $\mathit{inset}$ of both figures shows the corresponding local solution of the full system (\ref{eq:HPsc}) at $T=60$ for ($\mathit{a}$) $H$ and ($\mathit{b}$) $P$.}
    \label{fig:HhatPhat}
\end{figure*}
Figure \ref{fig:HhatPhat} shows the solution of $\hat{H}$ and $\hat{P}$ for $\gamma_{1,1}=\gamma_{1,2}=1$. Variations in the value, even sign, of the parameters $\gamma_{1,1}$ and $\gamma_{1,2}$ yield identical results to that presented in figure $\ref{fig:HhatPhat}$.

The solutions given in figure \ref{fig:HhatPhat} show excellent agreement locally with the solutions of the full system (\ref{eq:HPsc}), solved numerically and presented in figure \ref{fig:HbarPbar}, at large time, especially the solution found for $\hat{P}$ which is qualitatively identical to the solution of $\bar{P}$ at large times (the large time solution to the full system is shown in the inset of figure \ref{fig:HhatPhat}, for comparison).

Now, let us move further rightwards into positive $\xi$ by considering $\xi=\mathcal{D}\bar{X}$, where $\mathcal{D}\gg1$ and $\bar{X}$ is of order unity. We can infer expansions of $\hat{H}$ and $\hat{P}$ from their large positive $\xi$ asymptotes (\ref{eq:largepxi}-\ref{eq:largepxip}),
\begin{subequations}
\begin{equation}
\hat{H}= \mathcal{D}(\hat{H}_0(3\ln\mathcal{D})^{1/3}+\hat{H}_1(3\ln\mathcal{D})^{-2/3}+\cdots),
\end{equation}
\begin{equation}
\hat{P}= \mathcal{D}^{-1}(\hat{P}_0(3\ln\mathcal{D})^{-2/3}+\hat{P}_1(3\ln\mathcal{D})^{-5/3}+\cdots).
\end{equation}
\end{subequations}
To leading order, equation (\ref{eq:Htd}) then yields $\mathrm{d}^3\hat{H}_0/\mathrm{d}\bar{X}^3=0$ and matching requires $\hat{H}_0\sim\lambda_1\bar{X}$ as $\bar{X}\rightarrow0^+$. Hence 
\begin{equation}
    \hat{H}_0=\lambda_1\bar{X}+\mu_1\bar{X}^2,
\end{equation}
where $\mu_1$ is a positive constant. Now we can see that, at leading order for both $\hat{H}$ and $\hat{P}$, 
\refstepcounter{equation} \label{eq:HPlargeX}
$$
  \hat{H}\sim\mu_1\bar{X}^2, ~~~~~
  \hat{P}\sim0, ~~~~~ \mbox{as\ }\bar{X}\rightarrow\infty
  \eqno{(\theequation{\text{a},\text{b}})},
$$
which is a form resembling the far field condition (\ref{eq:HbPbff}). In light of (\ref{eq:xi}), (\ref{eq:HPexpH}) and (\ref{eq:HPlargeX}$\text{a}$), $H$ emerges as
\begin{equation}
    H\sim\mu_1\mathcal{D}^{-2}T^{2(1-\alpha)/3-\lambda}(X-cT^\alpha)^2
\end{equation}
just to the right of the local zone, which, as far as the $X^2$ requirement in (\ref{eq:HbPbff}$\text{a}$) is concerned, indicates that the constants must satisfy
\refstepcounter{equation}
$$
    \alpha=1-\frac{3\lambda}{2},~~~~~
    m=\frac{\lambda}{2};
    \eqno{(\theequation{\text{a},\text{b}})}
$$
hence any $\lambda\leq0$ works here. To the right of the above local region we have the balance $H_{TT}\sim0$, which when matching to (\ref{eq:HbPbff}$\text{a}$) at infinity gives the solution
\begin{equation} \label{eq:ltff}
    H \sim \frac{X^2}{2}-T,
\end{equation}
which in turn matches with the far-field solution.

The most acceptable looking solution to arise from this analysis would be if $\lambda=0$, resulting in $\alpha=1$ and $m=0$, with $\xi=X-cT$ and $H$ and $P$ depending on $\xi$ alone. This nonlinear travelling wave form is described fully by (\ref{eq:xi}-\ref{eq:ltff}), capturing the large time features of the full system (\ref{eq:HPsc}) successfully and confirming the absence of touchdown on this time scale.

The analysis here is a similar one to that seen in \citet{purvis2004} and is analogous to a finite-time break up formulation \cite{smith1988,smith2003}. A physical interpretation of this large time analysis of the large surface deformation system could well be the increased gliding extent observed in \citet{langley2020} for droplet impacts onto soft solids. Gliding occurs when the droplet does not make contact with the substrate at the kink of the dimple, and instead glides on a thin air layer. In \citet{langley2020} this is seen to occur more frequently and to a greater extent for more deformable surfaces. This could be interpreted as what is occurring in the numerical results for the reduced system in figure \ref{fig:HbarPbar}, with the increased time scales and horizontal extent of the kink being potential traits of droplet gliding in the absence of defects and localised wetting \citep{langley2017}.

\section{Conclusions} \label{conc}

A fluid-structure interaction model describing the pre-impact air cushioning behaviour of a droplet impacting a deformable surface has been developed. It rationally couples the thin film lubrication flow of the air to an approaching quasi-inviscid droplet approaching from above and a membrane-type model of the deformable surface below. Building on previous work by \citet{smith2003}, the model assumes a deformable surface deflection of the same order as the droplet free-surface deformation, which allows us to couple the surface deflection with the air film pressure and free-surface deflection. 

The deflection of the surface depends on the parameters describing the surface properties, which are incorporated into the membrane type deformable surface equation. These parameters correspond to the surface rigidity, tension, spring stiffness, mass density and damping. We considered two separate major cases of the surface equation. The first, a viscoelastic model, only considered the surface stiffness and the (viscous) damping. Numerical solutions to this system were presented and a number of conclusions were drawn. It was found that by lowering the magnitude of the surface parameters and increasing deformability, the approach to touchdown is considerably delayed, pressure buildup is decreased and more air is entrapped as touchdown is approached. For sufficiently low magnitude parameters, the pressure peak as touchdown is approached is a round one, as opposed to a sharp peak seen before on a rigid surface. The pressure peak is also seen to be located just behind the advancing cusp of the air film thickness, resulting in an increased extent of the air film as touchdown is approached. A numerical analysis was also conducted on the effect of variations of droplet radius and impact velocity on the size of entrapped air at impact, comparing a flat rigid surface to soft viscoelastic surfaces. It was found that the increase in entrapped air due to a soft viscoelastic surface was almost independent of droplet radius, on a logarithmic scale, while it is strongly dependent on the droplet velocity. For a flat rigid surface, increased droplet velocity results in decreased air entrapment, while for impact onto a soft solid this decrease in air entrapment is delayed and even halted for increased droplet velocity. A more general, flexible surface was then considered and broadly the same conclusions can be drawn here as for the viscoelastic surface case. A substantial difference however is that for a flexible surface, the pressure peak as touchdown is approached remains sharp, despite reductions in amplitude, for increased flexibility. It is shown that reductions in magnitude of each parameter corresponding to the flexural rigidity, tension, stiffness and mass density (for an undamped flexible surface) resulted in increased air entrapment. Qualitative connections of these findings were made to recent experimental work by \citet{howland2016} and \citet{langley2020}.

A case was then considered where we sought a solution in the limit of large surface deformations, for which the air film thickness is dominated by the surface deformation rather than by the free-surface deformation. This leads to a new system of governing equations which is similar to that of the flat rigid surface impact, albeit with a new pressure-shape relationship. This new system, solved computationally, showed results highlighting the significant increase in horizontal bubble extent and delay to touchdown. A large time analysis was found to give excellent agreement with the full numerical results and confirmed the apparent absence of touchdown (thus hinting at so-called gliding) for that particular system.

\begin{acknowledgements}
N.I.J.H. is grateful for the UCL Mechanical Engineering and Mathematics Studentship funding his PhD, during which this work was undertaken. N.I.J.H. is also very grateful for the use of the UCL Mechanical Engineering Minion High Performance Computer Cluster and associated support services. F.T.S. thanks EPSRC and UCL for continued support, including grants EP/R511638/1, GR/T11364/01, EP/G501831/1, EP/H501665/1, EP/K032208/1. M.K.T. acknowledges the funding from the European Union's Horizon 2020 Research and Innovation programme under the European Research Council (ERC) grant 714712 (NICEDROPS) and the Royal Society Wolfson Fellowship. The authors would like to thank Professor A. A. Korobkin for kindly pointing out to us, after completion of this work, the thesis of \citet{pegg2019}.
\end{acknowledgements}


%
%

%


\bibliography{bibliography}

\end{document}